\documentclass[aps,floatfix,showpacs,superscriptaddress,showkeys,preprint]
{revtex4}

\usepackage{graphicx,epsfig,epstopdf}
\usepackage{amssymb,amsmath,amsxtra,amsfonts}
\usepackage{bm}

\usepackage{longtable}
\usepackage{multirow}
\usepackage{booktabs}
\usepackage{array}
\usepackage{wrapfig}
\usepackage{color}

\usepackage{titlesec}
\titleformat{\section}{\large\bfseries}{\thesection}{1em}{}

\newcommand{\bea}{\begin{eqnarray}}
\newcommand{\ena}{\end{eqnarray}}
\newcommand{\be}{\begin{equation}}
\newcommand{\en}{\end{equation}}
\newcommand{\nn}{\nonumber\\}
\newcommand{\ed}{\end{document}} 
\newcommand{\Tr}{\mbox{\rm{tr}}}
\newcommand{\ord}{\mathcal{O}}

\newcommand{\bl}{\bigl}
\newcommand{\br}{\bigr}
\newcommand{\la}{\langle}
\newcommand{\ra}{\rangle}

\newcommand{\Jpsi}{\ensuremath{J\!/\!\psi}}

\begin{document}

\title{Decay $B_s\to \phi \ell^+ \ell^-$ in covariant quark model}

\author{S.~Dubni\v{c}ka}
\affiliation{Institute of Physics, Slovak Academy of Sciences, 
Bratislava, Slovakia} 

\author{A.Z.~Dubni\v{c}kov\'{a} }
\affiliation{Comenius University, Bratislava, Slovakia}

\author{A.~Issadykov}
\affiliation{
Joint Institute for Nuclear Research, Dubna, Russia}
\affiliation{Faculty of physics and technical sciences, 
L.N.Gumilyov Eurasian National University, 010008 Astana, 
Republic of Kazakhstan}

\author{M.A.~Ivanov}
\email{ivanovm@theor.jinr.ru}
\affiliation{
Joint Institute for Nuclear Research, Dubna, Russia}

\author{A.~Liptaj}
\email{andrej.liptaj@savba.sk}
\affiliation{Institute of Physics, Slovak Academy of Sciences, 
Bratislava, Slovakia} 

\author{S.K.~Sakhiyev} 
\affiliation{Faculty of physics and technical sciences, 
L.N.Gumilyov Eurasian National University, 010008 Astana, 
Republic of Kazakhstan}
\affiliation{Institute of Nuclear Physics, 050032 Almaty, 
Republic of Kazakhstan}

\begin{abstract}
Our article is devoted to the study of the rare $B_s\to \phi \ell^+\ell^-$~decay
where $\ell=\mu,\tau$. We compute the relevant form factors in the framework 
of the covariant quark model with infrared confinement in the full kinematical 
momentum transfer region. The calculated form factors are used to evaluate
branching fractions and polarization observables in the cascade decay
$B\to \phi(\to K^+K^-)\ell^+\ell^-$. We compare the obtained results with 
available experimental data and the results from other theoretical approaches.


\end{abstract}

\maketitle

\section{Introduction}
\label{sec:intro}

The transition $b\to s\ell^+\ell^-$ mediated by Flavor-Changing Neutral
Current (FCNC) is one of the key point in the Standard Model (SM)
which allows one to look for the possible manifestation of
New Physics (NP). The physical processes induced by this transition
are currently studied in great details at the LHC. 
The most popular and well--analyzed among them are the rare B-meson decays
$B\to K^\ast(\to K\pi)\mu^+\mu^-$ and  $B_s\to \phi(\to K^+K^-)\mu^+\mu^-$.
The decay $\Lambda_b \to \Lambda(\to p\pi)\,\ell^+\ell^-$
can be considered to be a welcome complement to the above decay channels.

The  LHCb Collaboration \cite{Aaij:2013qta} reported a measurement of 
form-factor independent angular observables in the decay 
$ B\to K^\ast\mu^+\mu^-$.
One observable was found to be in  disagreement with the SM 
on the level of 3.7~$\sigma$.

The improved measurements of the isospin asymmetries and branching fractions
for $B\to K\mu^+\mu^-$ and $ B\to K^\ast\mu^+\mu^-$ decays were reported in
\cite{Aaij:2014pli}. The isospin asymmetries were consistent with the SM,
whereas some branching fractions were found to be slightly lower
than the theoretical predictions.

An angular analysis and a measurement of the differential branching fraction 
of the decay $ B^0_s\to \phi \mu^+\mu^-$ were presented in \cite{Aaij:2015esa}.
The results of the angular analysis are consistent with the SM.
However, the differential branching fraction in one bin was found
to be more than 3~$\sigma$  below the SM predictions.

The observed discrepancies (sometimes called ``$b\to s\ell\ell$ anomalies'')
have generated a plenty of theoretical studies
\cite{Descotes-Genon:2015uva}-\cite{Kang:2013jaa} involving
the various scenarios of NP and analysis of the uncertainties  
from hadronic contributions. 
The form factors obtained from unquenched lattice QCD \cite{Horgan:2013hoa}
were used in \cite{Horgan:2013pva,Horgan:2015vla} to calculate the differential
 branching fractions of the decays  
$ B\to K^\ast\mu^+\mu^-$ and  $B_s\to \phi \mu^+\mu^-$.

In this paper we calculate all form factors which appear
in the $B_s\to\phi$~transition by using the covariant quark model.
The expressions for the Wilson coefficients $C_7$ and $C_9$  
are taken on the two-loop level of accuracy by using the results
obtained in Refs.~\cite{Asatryan:2001zw,Greub:2008cy}.
Then we evaluate the branching fraction, the forward-backward 
asymmetry and the so-called optimized observables in the cascade decay
$B_s\to \phi(\to K^+K^-)\mu^+\mu^-$.  We compare our results 
with the recent experimental data reported in Ref.~\cite{Aaij:2015esa}
for various $q^2$-bins.

\section{Model}
\label{sec:model}

The covariant confined quark model 
\cite{Efimov:1988yd,Efimov:1993ei,Faessler:2002ut,Branz:2009cd}
is an effective quantum field approach to hadronic interactions based on 
an interaction Lagrangian of hadrons interacting with their constituent quarks.
The value of the coupling constant follows form the compositeness 
condition~$Z_H=0$, where $Z_H$ is the wave function renormalization constant of the hadron. Matrix elements of the physical processes are generated by a set 
of quark loop diagrams according to the $1/N_c$ expansion. The ultraviolet 
divergences of the quark loops are regularized by including vertex functions 
for the hadron-quark vertices. These functions also describe finite size 
effects related to the non-pointlike hadrons. The quark 
confinement \cite{Branz:2009cd} is built-in through an infrared cutoff on 
the upper limit of the scale integration 
to avoid the appearance of singularities in matrix elements. 
The infrared cutoff parameter $\lambda$ is universal for all processes. 
The  covariant confined quark model has limited number of parameters: the light and heavy constituent quark masses, 
the size  parameters which describe the size of the distribution 
of the constituent quarks inside the hadron and 
the infrared cutoff parameter $\lambda$. They are determined by a
fit to available experimental data.

Let us start with the effective Lagrangian describing
the transition of a meson $M(q_1\bar q_2)$ to its constituent
quarks $q_1$ and $\bar q_2 $ 
\bea
{\mathcal L}_{\rm int}(x) &=& g_M M(x)\cdot J_M(x) + {\rm h.c.},
\nn
J_M(x) &=& \int\!\! dx_1 \!\!\int\!\!
dx_2 F_M (x,x_1,x_2)\bar q_2(x_2)\Gamma_M q_1(x_1) 
\ena
with $\Gamma_M$ a Dirac matrix which projects onto the spin quantum 
number of the meson field $M(x)$. The vertex function $F_M$  characterizes 
the finite size of the meson. Translational invariance requires the 
function $F_M$ to fulfill the identity $F_M(x+a,x_1+a,x_2+a)=F_M(x,x_1,x_2)$ for
any four-vector $a$. A specific form for the  vertex function is adopted
\be
F_M(x,x_1,x_2)=\delta(x - w_1 x_1 - w_2 x_2) \Phi_M((x_1-x_2)^2),
\label{eq:vertex}
\en
where $\Phi_M$ is the correlation function of the two constituent quarks
with masses $m_{q_1}$, $m_{q_2}$ and the mass ratios
$w_i = m_{q_i}/(m_{q_1}+m_{q_2})$.

A simple Gaussian form of the vertex function $\bar \Phi_M(-\,k^2)$ is selected
\be
\bar \Phi_M(-\,k^2) 
= \exp\left(k^2/\Lambda_M^2\right)
\label{eq:Gauss}
\en
with the parameter $\Lambda_M$ linked to the size of the meson. The minus sign 
in the argument is chosen to indicate that we are working in the Minkowski 
space. Since $k^2$ turns into $-\,k_E^2$ in the Euclidean space, 
the form (\ref{eq:Gauss}) has the appropriate fall-off behavior in 
the Euclidean region. Any choice for  $\Phi_M$ is appropriate
as long as it falls off sufficiently fast in the ultraviolet region of
the Euclidean space to render the corresponding Feynman diagrams ultraviolet 
finite. We choose a Gaussian form for calculational convenience.

The fermion propagators for the quarks are given by
\be
S_i(k)=\frac{1}{m_{q_i}-\not\! k}
\label{eq:prop}
\en
with an effective constituent quark mass $m_{q_i}$. 

The so-called {\it compositeness condition} 
\cite{Weinberg:1962hj, Salam:1962ap, Hayashi:1967hk, Efimov:1993ei} is used 
to determine the value of the coupling constants $g_M$. 
It means that the renormalization constant $Z_M$ of the elementary meson 
field $M(x)$ is set to zero, i.e.,
\be
Z_M \, = \, 1 - \, \frac{3g^2_M}{4\pi^2} \,\bar\Pi'_M(m^2_M) \, = \, 0,
\label{eq:Z=0}
\en
where $\bar\Pi^\prime_M$ is the derivative of the meson mass operator.
Its physical meaning in Eq.~(\ref{eq:Z=0}) becomes clear when interpreted as 
the matrix element between the physical and the corresponding bare state: 
$Z_M=0$ implies that the physical state does not contain the bare state and 
is appropriately described as a bound state. The interaction makes 
the physical particle dressed, i.e. its mass and wave function have to be 
renormalized. The condition $Z_M=0$ also effectively excludes the constituent 
degrees of freedom from the space of physical states. It thereby guarantees 
the absence of double counting for the physical observable under consideration,
the constituents exist only in  virtual states. The tree-level diagram together
with the diagrams containing self-energy insertions into the external legs 
(i.e. the tree-level diagram times $Z_M -1$) give
a common factor $Z_M$  which is equal to zero.

The mass functions for the pseudoscalar meson (spin $S=0$)
and vector meson (spin $S=1$) are defined as
\bea
\Pi_{P}(x-y) &=& +\,i\,\la T\bl\{J_P(x)J_P(y) \br\} \ra_0 ,
\label{eq:S=0}\\[2ex]
\Pi^{\mu\nu}_{V}(x-y) &=& -\,i\,\la T\bl\{J^\mu_V(x)J^\nu_V(y) \br\} \ra_0 .
\label{eq:S=1}
\ena
By using the Fourier transforms of the vertex functions in Eq.~(\ref{eq:Gauss})
and quark propagators in Eq.(\ref{eq:prop}) one can easily find
the Fourier transforms of the mass functions
\bea
\tilde\Pi_{P}(p^2) &=& N_c\int\frac{d^4k}{(2\pi)^4i} \tilde\Phi^2_P(-k^2)
\Tr\Big(\gamma^5 S_1(k+w_1 p)\gamma^5 S_2(k-w_2 p)\Big),
\label{eq:massP-1}\\[2ex]
\tilde\Pi^{\mu\nu}_{V}(p) &=& N_c\int\frac{d^4k}{(2\pi)^4i} \tilde\Phi^2_V(-k^2)
\Tr\Big(\gamma^\mu S_1(k+w_1 p)\gamma^\nu S_2(k-w_2 p)\Big)
\nn
&=& g^{\mu\nu} \tilde\Pi_{V}(p^2) + p^\mu p^\nu  \tilde\Pi_{V}^\parallel(p^2)
\label{eq:massV-1}
\ena
where $N_c=3$ is the number of colors. Due to the transversality of the
vector field the second term in Eq.~(\ref{eq:massV-1}) is irrelevant
in our consideration. The first term in Eq.~(\ref{eq:massV-1}) may be
picked out as
\be
 \tilde\Pi_{V}(p^2) = \frac13 \bl(g_{\mu\nu}-\frac{p_\mu p_\nu}{p^2} \br)
                       \tilde\Pi^{\mu\nu}_{V}(p).
\label{eq:massV-2}
\en

The loop integrations in Eqs.~(\ref{eq:massP-1}) and 
~(\ref{eq:massV-1}) are done with the help of the Fock-Schwinger 
representation of the quark propagator
\bea
S_q (k + p) &=& \frac{1}{ m_q-\not\! k- \not\! p } 
=  \frac{m_q + \not\! k +  \not\! p}{m^2_q - (k+ p)^2}
\nn
&=& (m_q + \not\! k +  \not\! p)\int\limits_0^\infty \!\!d\alpha\, 
e^{-\alpha [m_q^2-(k+p)^2]}\,,
\label{eq:Fock}
\ena
where $k$ is the loop momentum and $p$ is the external momentum.
As described later on, the use of the Fock-Schwinger representation allows 
one to do tensor loop integrals in a very efficient way since one can
convert loop momenta into derivatives of the exponential function. 

All loop integrations are performed in Euclidean space. 
The transition from Minkowski space to Euclidean space is performed
by using the Wick rotation
\be
k_0=e^{i\frac{\pi}{2}}k_4=ik_4
\label{eq:Wick}
\en
so that $k^2=k_0^2-\vec{k}^2=-k_4^2-\vec{k}^2=-k_E^2 \leq 0.$
Simultaneously one has to rotate all external momenta, i.e.
 $p_0 \to ip_4$ so that $p^2=-p_E^2 \leq 0$.
Then the quadratic form in Eq.~(\ref{eq:Fock}) becomes positive-definite

\[
m^2_q-(k+p)^2=m^2_q + (k_E+p_E)^2>0,
\]
and the integral over $\alpha$ is absolutely convergent.
We will keep the Minkowski notation to avoid
excessive relabeling. We simply imply that
 $k^2 \leq 0$ and $p^2 \leq 0$.

Collecting the representations for the vertex functions
and quark propagators given by Eqs.~(\ref{eq:Gauss})
and (\ref{eq:Fock}), respectively, one can perform the Gaussian
integration in the expressions for  the matrix elements
in Eqs.~(\ref{eq:massP-1}) and ~(\ref{eq:massV-1}). 
The exponent has the form $ak^2+2kr+z_0$, where $r=b\,p$. 
Using the following properties 
\be
\begin{aligned}
    k^\mu\, \exp(ak^2+2kr+z_0) &=\frac{1}{2}\frac{\partial }
{\partial r_\mu }\exp(ak^2+2kr+z_0)\\
    k^\mu k^\nu\, \exp(ak^2+2kr+z_0) &=
      \frac{1}{2}\frac{\partial }{\partial r_\mu } 
      \frac{1}{2} \frac{\partial }{\partial r_\nu }         \exp(ak^2+2kr+z_0)
\\
\text{etc.}&
\end{aligned}
\label{eq:change-to-r}\,
\en
one can replace
$\not\! k $ by $ {\not\! \partial}_r 
= \gamma^\mu\frac{\partial}{\partial r_\mu}$
which allows one to exchange the tensor integrations
for a differentiation of the Gaussian exponent $e^{-r^2/a}$
which appears after integration over loop momentum. 
The $r$-dependent Gaussian exponent $e^{-r^2/a}$ can be moved to the left 
through the differential operator $\not\! \partial_r$ by using the following 
properties
\bea
\frac{\partial}{\partial r_\mu}\,e^{-r^2/a} &=& e^{-r^2/a}
\left[-\frac{2r^\mu}{a}+\frac{\partial}{\partial r_\mu}\right]\,,
\nn[1.2ex]
\frac{\partial}{\partial r_\mu}\,
\frac{\partial}{\partial r_\nu}\,e^{-r^2/a} &=& e^{-r^2/a}
\left[-\frac{2r^\mu}{a}+\frac{\partial}{\partial r_\mu}\right]\cdot
\left[-\frac{2r^\nu}{a}+\frac{\partial}{\partial r_\nu}\right]\,,
\nn[1.2ex]
\text{etc.}&&
\label{eq:dif}
\ena
Finally, one has to move the derivatives to the right by using
the commutation relation
\be
\left[\frac{\partial}{\partial r_\mu},r^\nu \right]
=g^{\mu\nu}\,.
\label{eq:comrel}
\en
The last step has been done by using a FORM code which
works for any numbers of loops and propagators.
In the remaining integrals over the Fock-Schwinger parameters 
$0\le \alpha_i<\infty$
we introduce an additional integration which converts the set of 
Fock-Schwinger parameters into a simplex. We use the transformation:
\be
\prod\limits_{i=1}^n\int\limits_0^{\infty} 
\!\! d\alpha_i f(\alpha_1,\ldots,\alpha_n)
=\int\limits_0^{\infty} \!\! dtt^{n-1}
\prod\limits_{i=1}^n \int\!\!d\alpha_i 
\delta\left(1-\sum\limits_{i=1}^n\alpha_i\right)
  f(t\alpha_1,\ldots,t\alpha_n)
\label{eq:simplex}  
\en

Finally, one finds
\bea
\tilde\Pi_M(p^2)&=& \frac{3}{4\pi^2}\int\limits_0^\infty \!\! 
\frac{dt\,t}{a_M^2} \int\limits_0^1\!\!d\alpha\,
e^{-t\,z_0 + z_M}\,
\Big\{
  \frac{n_M}{a_M} + m_{q_1}m_{q_2} 
+ \bl(w_1 - \frac{b}{a_M}\br) \bl(w_2 + \frac{b}{a_M}\br)p^2
\Big\}
\label{eq:mass_fin}\\[2ex]
z_0 &=&  \alpha m^2_{q_1} +(1-\alpha)m^2_{q_2} - \alpha(1-\alpha) p^2 ,
\qquad z_M  = \frac{2s_Mt}{2s_M+t} (\alpha-w_2)^2 p^2 ,
\nn[2ex] 
a_M &=& 2s_M+t\, , \qquad b = (\alpha-w_2)t\,. 
\nonumber
\ena
Here $n_M=2$ for a pseudoscalar and $n_M =1$ for vector particle.  
The parameter $s_M$
is related to  the size parameter  $\Lambda_M$  as $s_M=1/\Lambda^2_M$.

The integral over ``t'' is well-defined and convergent 
below the threshold  $p^2< (m_{q_1} + m_{q_2})^2$.  
The convergence of the integral above  threshold  
$p^2\ge (m_{q_1} + m_{q_2})^2$ is guaranteed by the addition of a small imaginary part 
to the quark mass, i.e. $m_q\to m_q - i\epsilon, \quad \epsilon>0$
in the quark propagator. It allows one to rotate
the integration variable ``t'' to the imaginary axis $t\to i t$. 
As a result the integral  becomes convergent but obtains an imaginary part 
corresponding to quark pair production.

However, by cutting the scale integration at the upper limit, which corresponds to the introduction of an infrared cutoff
\be
\int\limits_0^\infty dt (\ldots) \to \int\limits_0^{1/\lambda^2} dt (\ldots)
\label{eq:conf}
\en
one can remove all possible thresholds present in the initial quark
diagram~\cite{Branz:2009cd}. Thus the infrared cutoff parameter 
$\lambda$ effectively guarantees the confinement of quarks within hadrons. 
This method is quite general and can be used for diagrams with an arbitrary 
number of loops and propagators. 

\section{Form factors of the \boldmath{$B_s\to\phi$} transition}
\label{sec:form_factors}

\begin{figure*}[htbp]
\begin{center}
\includegraphics[width=0.90\textwidth]{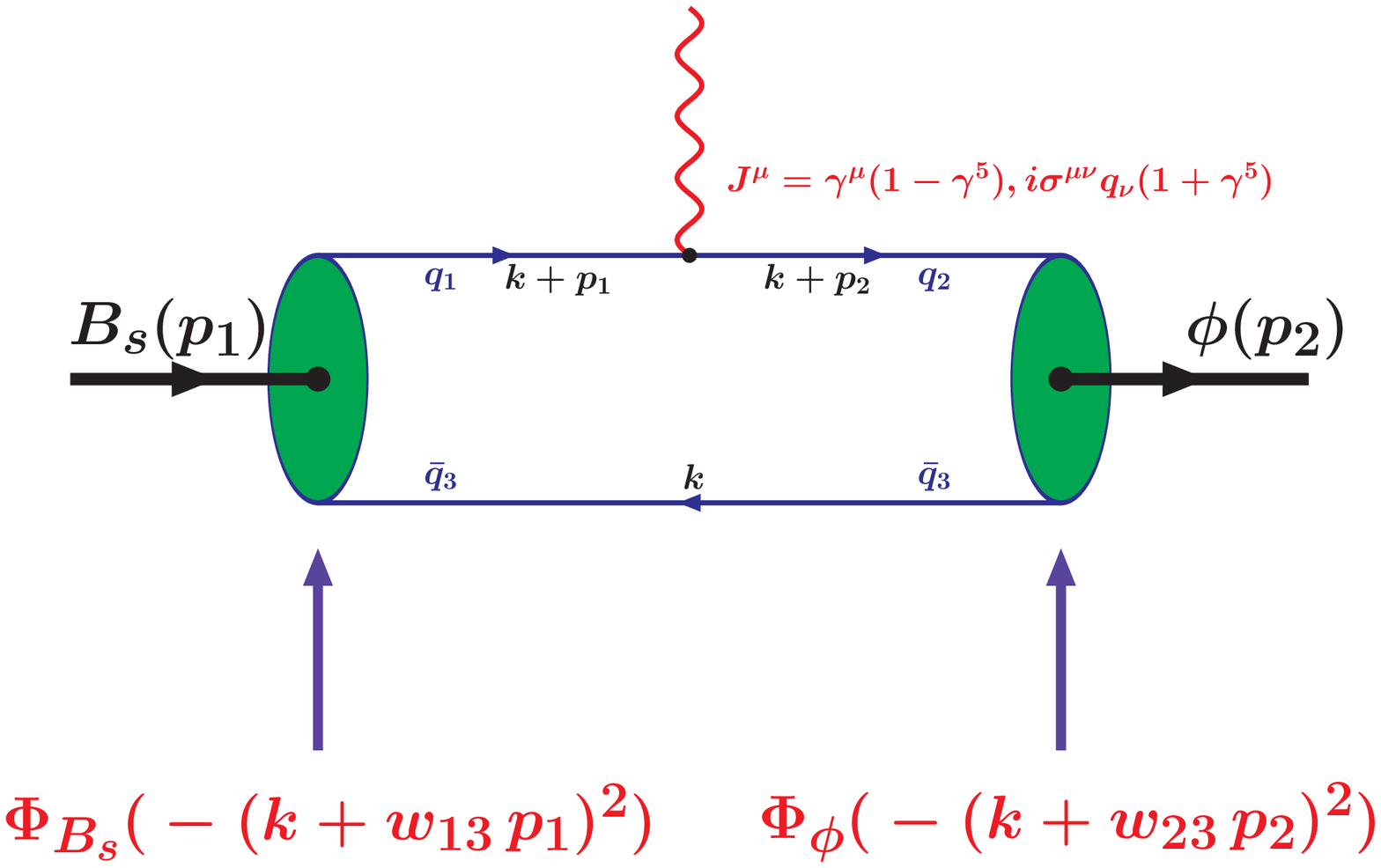} 
\caption{Diagrammatic representation of the matrix elements
describing $B_s\to \phi$ transitions. Identification of quarks:
$q_1=b$, $q_2=q_3=s$, $w_{13}=m_s/(m_b+m_s)$ and $w_{23}=1/2$. }
\label{diag_BsPhi}
\end{center}
\end{figure*}

The Feynman diagram describing the $B_s\to\phi$ transition
in the framework of our covariant quark model is depicted
in Fig.~\ref{diag_BsPhi}.  The matrix element is expressed
through dimensionless form factors \cite{Dubnicka:2015iwg,Ivanov:2011aa}:
\bea
&&
\langle 
\phi(p_2,\epsilon_2)\,
|\,\bar s\, O^{\,\mu}\,b\, |\,B_{s}(p_1)
\rangle 
\,=\,
\nn
&=&
N_c\, g_{B_s}\,g_\phi \!\! \int\!\! \frac{d^4k}{ (2\pi)^4 i}\, 
\widetilde\Phi_{B_s}\Big(-(k+w_{13} p_1)^2\Big)\,
\widetilde\Phi_\phi\Big(-(k+w_{23} p_2)^2\Big)
\nn
&\times&
{\rm tr} \biggl[ 
O^{\,\mu} \,S_b(k+p_1)\,\gamma^5\, S_s(k) \not\!\epsilon_2^{\,\,\dagger} \,
S_s(k+p_2)\, \biggr]
\nn
 & = &
\frac{\epsilon^{\,\dagger}_{\,\nu}}{m_1+m_2}\,
\Big( - g^{\mu\nu}\,P\cdot q\,A_0(q^2) + P^{\,\mu}\,P^{\,\nu}\,A_+(q^2)
       + q^{\,\mu}\,P^{\,\nu}\,A_-(q^2)
\nn 
&& + i\,\varepsilon^{\mu\nu\alpha\beta}\,P_\alpha\,q_\beta\,V(q^2)\Big),
\label{eq:PV}
\ena
\bea
&&
\langle 
\phi(p_2,\epsilon_2)\,
|\,\bar s\, (\sigma^{\,\mu\nu}q_\nu(1+\gamma^5))\,b\, |\,B_{s}(p_1)
\rangle 
\,=\,
\nn
&=&
N_c\, g_{B_s}\,g_\phi \!\! \int\!\! \frac{d^4k}{ (2\pi)^4 i}\, 
\widetilde\Phi_{B_s}\Big(-(k+w_{13} p_1)^2\Big)\,
\widetilde\Phi_\phi\Big(-(k+w_{23} p_2)^2\Big)
\nn
&\times&
{\rm tr} \biggl[ 
(\sigma^{\,\mu\nu}q_\nu(1+\gamma^5))
\,S_b(k+p_1)\,\gamma^5\, S_s(k) \not\!\epsilon_2^{\,\,\dagger} \,S_s(k+p_2)\, 
\biggr]
\nn
 & = &
\epsilon^{\,\dagger}_{\,\nu}\,
\Big( - (g^{\mu\nu}-q^{\,\mu}q^{\,\nu}/q^2)\,P\cdot q\,a_0(q^2) 
       + (P^{\,\mu}\,P^{\,\nu}-q^{\,\mu}\,P^{\,\nu}\,P\cdot q/q^2)\,a_+(q^2)
\nn
&&
+ i\,\varepsilon^{\mu\nu\alpha\beta}\,P_\alpha\,q_\beta\,g(q^2)\Big).
\label{eq:PVT}
\ena

Here, $P=p_1+p_2$, $q=p_1-p_2$, $\epsilon_2^\dagger\cdot p_2=0$,
$p_1^2=m_1^2\equiv m^2_{B_s}$, $p_2^2=m_2^2\equiv m^2_\phi$ and
the weak matrix  $O^{\,\mu} = \gamma^{\,\mu}(1-\gamma^5)$.
Since there are three quarks involved in these
processes, we introduce the notation with two subscripts
$w_{ij}=m_{q_j}/(m_{q_i}+m_{q_j})$ $(i,j=1,2,3)$ so that $w_{ij}+w_{ji}=1$. 
The form factors defined in Eq.\,(\ref{eq:PVT}) satisfy the physical 
requirement $a_0(0)=a_+(0)$, which ensures that no kinematic singularity 
appears in the matrix element at $q^2=0 ~ \mathrm{GeV^2}$.  

Herein we use the updated values of the model parameters 
\cite{Dubnicka:2015iwg} which are shown in Eq.~(\ref{eq:fit}).
\be
\def\arraystretch{1.5}
\begin{array}{ccccc|ccc}
     m_{u/d}        &      m_s        &      m_c       &     m_b & \lambda  &   
 \Lambda_{B_s} & \Lambda_\phi
\\\hline
 \ \ 0.241\ \   &  \ \ 0.428\ \   &  \ \ 1.67\ \   &  \ \ 5.05\ \   & 
\ \ 0.181\ \   &\ \ 2.05\ \   &\ \ 0.88\ \   & \ {\rm GeV} 
\end{array}
\label{eq:fit}
\en

\begin{figure*}[htbp]
\begin{center}
\epsfig{figure=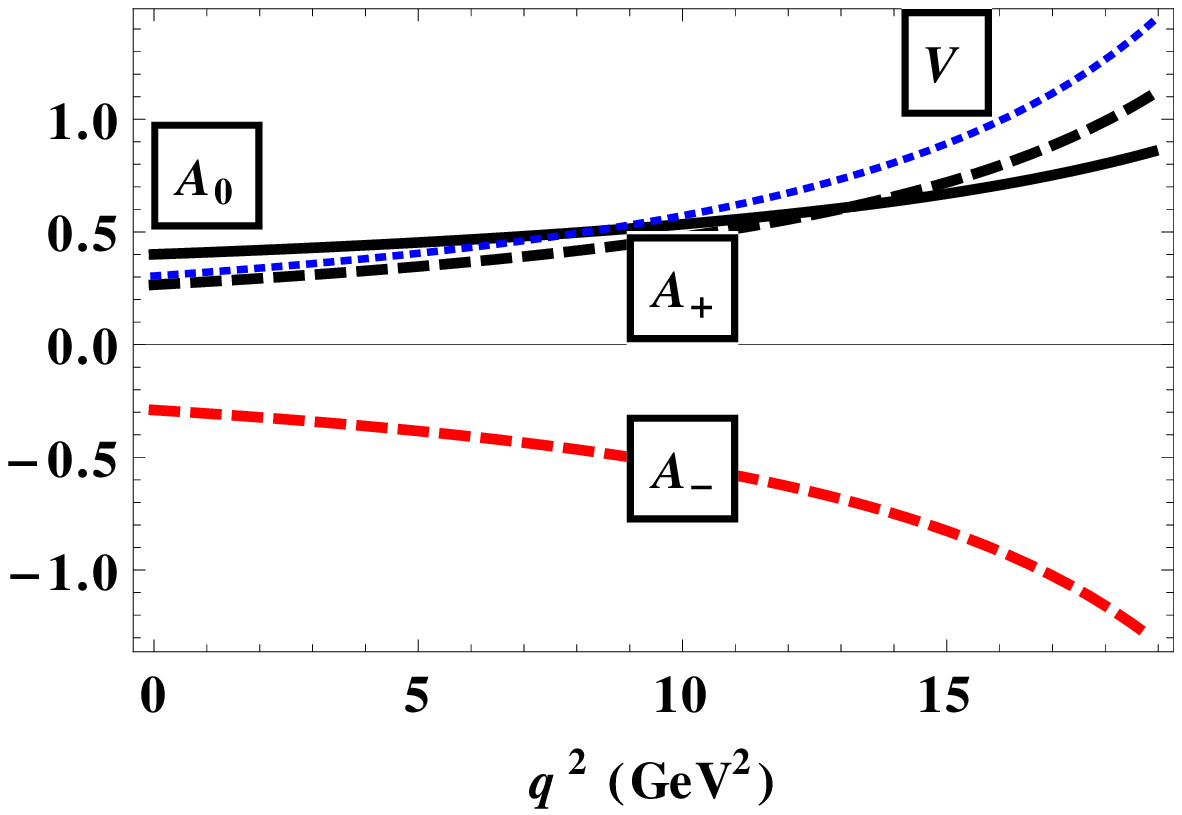,scale=1.15}
\hspace*{.25cm}
\epsfig{figure=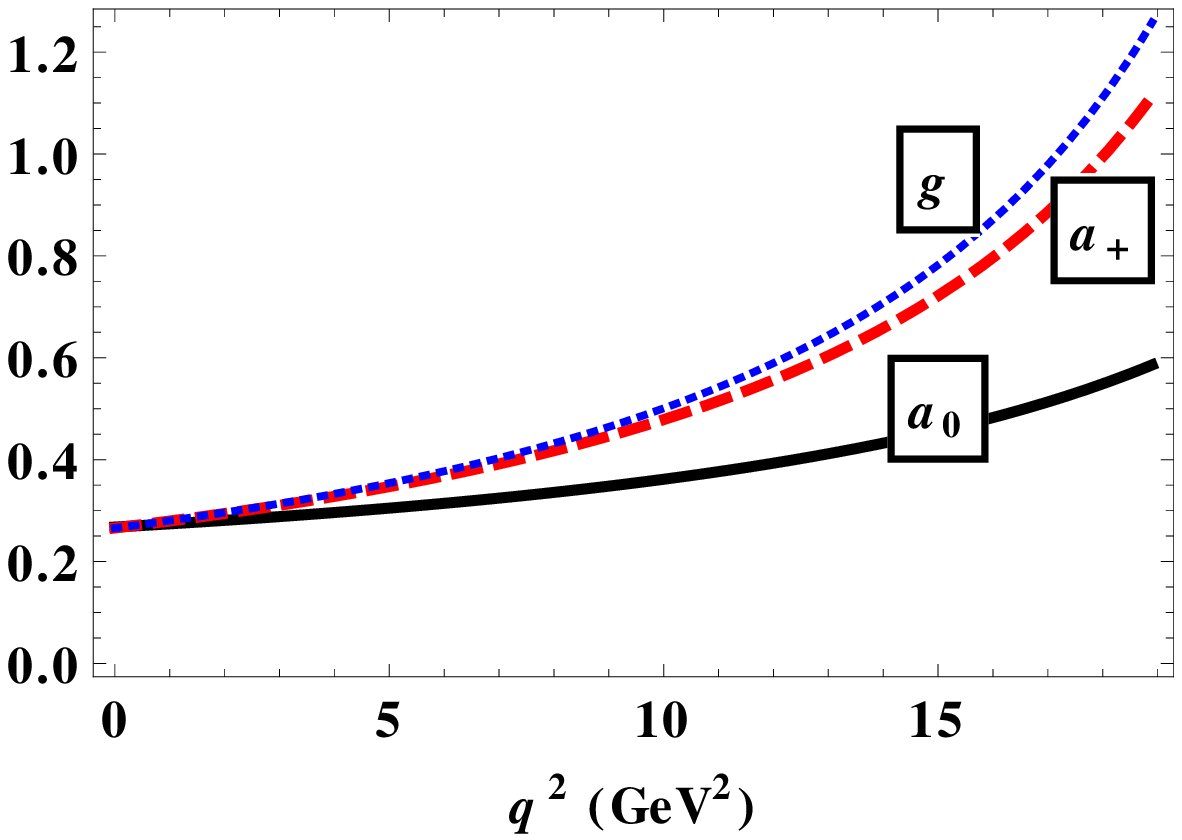,scale=1.08}
\caption{The  $q^{2}$-dependence of the vector and axial form factors
(upper plot) and tensor form factors  (lower plot).
\label{fig:BsPhi-FF}
}
\end{center}
\end{figure*}

Performing the loop integration in Eqs.~(\ref{eq:PV}) and (\ref{eq:PVT})
in a manner described in the previous section,
one can obtain the form factors in the form of three-fold integrals 
which are calculated numerically by using the FORTRAN code with NAG library.
The form factors are calculated in the full kinematical region of 
momentum transfer squared. 
The curves are depicted  in Fig.~\ref{fig:BsPhi-FF}.

The results of our numerical calculations are with high accuracy approximated 
by the parametrization
\be
F(q^2)=\frac{F(0)}{1-a s+b s^2}\,, \qquad s=\frac{q^2}{m_1^2}\,,
\label{eq:ff_approx}
\en
the relative error is less than 1$\%$.
The values of $F(0)$, $a$, and $b$ are listed  in Table~\ref{tab:apprff}.
\begin{table}[ht]
\caption{Parameters for the approximated form factors
in Eq.~(\ref{eq:ff_approx}).} 
\begin{center}
\begin{tabular}{c|rrrc|rrr}
\hline
&\qquad $A_0$ \qquad &\qquad $A_+$ \qquad  &\qquad $A_-$ \qquad & \qquad $V$ 
\qquad \quad 
&\qquad $a_0$ \qquad &\qquad $a_+$ \qquad &\qquad $g$ \qquad \\
\hline
$F(0)$ &  0.40   & 0.27 & $-0.29$ & 0.31 & 0.27    & 0.27 & 0.27 \\
$a$    &  0.62   & 1.41 & 1.48    & 1.51 & 0.66    & 1.41 & 1.52 \\
$b$    & $-0.30$ & 0.38 & 0.45    & 0.47 & $-0.26$ & 0.39 & 0.49 \\
\hline
\end{tabular}
\label{tab:apprff} 
\end{center}
\end{table}

For reference it is useful to relate the above form factors  
to those used, e.g., in Ref.\,\cite{Khodjamirian:2006st} 
(we denote them by the superscript $^c$).
The relations read 
\bea
A_0 &=& \frac{m_1 + m_2}{m_1 - m_2}\,A_1^c\,, \qquad 
A_+ = A_2^c\,,
\nn
A_- &=&  \frac{2m_2(m_1+m_2)}{q^2}\,(A_3^c - A_0^c)\,, \qquad
V = V^c\,, 
\nn[1.2ex]
a_0 &=& T_2^c\,, \qquad g = T_1^c\,, \qquad
a_+  =  T_2^c + \frac{q^2}{m_1^2-m_2^2}\,T_3^c\,.
\label{eq:new-ff}
\ena
We note in addition that the form factors (\ref{eq:new-ff}) satisfy
the constraints
\bea
 A_0^c(0) &=& A_3^c(0) 
\nn
2m_2A_3^c(q^2) &=& (m_1+m_2) A_1^c(q^2) -(m_1-m_2) A_2^c(q^2)\,.
\ena

Since $a_0(0)=a_+(0)=g(0)$ we display in Table~\ref{tab:ff-comparison}
the form factors
$A_0^c(0)=(m_1-m_2)[A_0(0)-A_{+}(0)]/(2m_2)$,
$A_1^c(0)=A_0(0)(m_1-m_2)/(m_1+m_2)$,
$A_2^c(0)=A_+(0)$,
$T_1^c(0)=g(0)$ and 
$T_3^c(0)=\lim_{\,q^2 \to 0}(m_1^2-m_2^2)(a_{+}-a_0)/q^2$
obtained in our model and compare them with
those from other approaches.

\begin{table}
\caption{The form factors at  maximum recoil $q^2=0$.}
\label{tab:ff-comparison}
\begin{ruledtabular}
\begin{tabular}{cllllll}
\toprule
     & $V^c(0)$ & $A_0^c(0)$ &$A_1^c(0)$ &$A_2^c(0)$ &$T_1^c(0)$ &$T_3^c(0)$ \\
\hline

This work & $0.31\pm 0.03$ & $0.28\pm 0.03$
 & $0.27\pm 0.03$ & $0.27\pm 0.03$ & $0.27\pm 0.03$ & $0.18\pm 0.02 $\\

Ref.~\cite{Ivanov:2011aa}& 0.32 &  & 0.29 & 0.28 & 0.28 & \\

Ref.~\cite{Ball:2004rg} &0.434$\pm$0.035 & $0.474\pm0.037$ & 0.311$\pm$0.029 
 & 0.234$\pm$0.028 & 0.349$\pm$0.033& $0.175\pm0.018$\\

Ref.~\cite{Faustov:2013pca} & $0.406\pm 0.020$ & $0.322\pm 0.016$ 
 & $0.320\pm 0.016$  & $0.318\pm 0.016$ & $0.275\pm0.014$& $0.133\pm 0.006$\\

Ref.~\cite{Yilmaz:2008pa}& 0.43& 0.38 &0.30  & 0.26& 0.35& 0.25\\

Ref.~\cite{Ali:2007ff} &$0.25\pm0.05$& $0.30\pm 0.05$ &$0.19\pm0.04$&&&\\

Ref.~\cite{Melikhov:2000yu} &$0.44$  & 0.42 & 0.34 &0.31&0.38& 0.26\\

Ref.~\cite{Li:2009tx} &$0.26\pm0.07$& $0.31\pm 0.07$ & $0.18^{+0.06}_{-0.05}$
 & $0.12\pm0.03$& $0.23^{+0.06}_{-0.05}$& $0.19\pm 0.05$\\

Ref.~\cite{Lu:2007sg}  &0.329 & 0.279 &0.232 &0.210&0.276& 0.170\\

Ref.~\cite{Wu:2006rd} &$0.339\pm0.017$&  &$0.271\pm0.014$&$0.212\pm0.011$
 &$0.299\pm0.016$& $0.191 \pm 0.010$\\
\end{tabular}
\end{ruledtabular}
\end{table}

\section{Effective Hamiltonian}
\label{sec:hamiltonian}

The rare decay $b \to s \ell^+ \ell^-$ is described in terms of 
the effective Hamiltonian \cite{Buchalla:1995vs}:  
\be
{\mathcal H}_{\rm eff} = - \frac{4G_F}{\sqrt{2}} \lambda_t   
              \sum_{i=1}^{10} C_i(\mu)  \ord_i(\mu) ,
\label{eq:effHam}
\en
where $C_i(\mu)$ and $\ord_i(\mu)$ are the Wilson coefficients and local 
operators, respectively. $\lambda_t~=~|V_{tb}V_{ts}^\ast|$ is the product 
of Cabibbo-Kobayashi-Maskawa (CKM) matrix elements. Note that we drop small corrections
 proportional to  $\lambda_u = |V_{ub}V_{us}^\ast|$.
The standard set \cite{Buchalla:1995vs} of local operators 
obtained within the SM for~$b \to s l^+ l^-$ transition is written as 

\bea 
\begin{array}{ll} 
\ord_1     =  (\bar{s}_{a_1}\gamma^\mu P_L c_{a_2})
              (\bar{c}_{a_2}\gamma_\mu P_L b_{a_1}),                   &
\ord_2     =  (\bar{s}\gamma^\mu P_L c)  (\bar{c}\gamma_\mu P_L b),   
\\[2ex]
\ord_3     =  (\bar{s}\gamma^\mu P_L  b) \sum_q(\bar{q}\gamma_\mu P_L q),  &
\ord_4     =  (\bar{s}_{a_1}\gamma^\mu P_L  b_{a_2}) 
              \sum_q (\bar{q}_{a_2}\gamma_\mu P_L q_{a_1}),
\\[2ex]
\ord_5     =  (\bar{s}\gamma^\mu P_L b)
              \sum_q(\bar{q}\gamma_\mu P_R q),            &
\ord_6     =  (\bar{s}_{a_1}\gamma^\mu P_L b_{a_2 }) 
              \sum_q  (\bar{q}_{a_2} \gamma_\mu P_R q_{a_1}),               
\\[2ex]
\ord_7     =  \frac{e}{16\pi^2} \bar m_b\, 
              (\bar{s} \sigma^{\mu\nu} P_R b) F_{\mu\nu},       &
\ord_8    =  \frac{g}{16\pi^2} \bar m_b\, 
              (\bar{s}_{a_1} \sigma^{\mu\nu} P_R {\bf T}_{a_1a_2} b_{a_2}) 
              {\bf G}_{\mu\nu},            
\\[2ex]
\ord_9     = \frac{e^2}{16\pi^2}        
             (\bar{s} \gamma^\mu P_L b) (\bar\ell\gamma_\mu \ell),     &
\ord_{10}  = \frac{e^2}{16\pi^2} 
             (\bar{s} \gamma^\mu P_L b)  (\bar\ell\gamma_\mu\gamma_5 \ell), 
\end{array}
\label{eq:operators}
\ena
where ${\bf G}_{\mu\nu}$ and $F_{\mu\nu}$ are the gluon and photon 
field strengths, respectively; ${\bf T}_{a_1a_2}$ are the generators of 
the $SU(3)$ color group; $a_1$ and $a_2$ denote color indices 
(they are omitted in the color-singlet currents).
The chirality projection operators are 
$P_{L,R} = (1 \mp \gamma_5)/2$ and $\mu$ is a renormalization scale.
$\ord_{1,2}$ are current-current operators, 
$\ord_{3-6}$ are QCD penguin operators,  $\ord_{7,8}$ are "magnetic  
penguin" operators, and $\ord_{9,10}$ are semileptonic electroweak penguin operators. We denote the QCD quark masses by the bar symbol
to distinguish them from the constituent quark masses used in the model. 

By using the effective Hamiltonian defined by Eq.~(\ref{eq:effHam})
one can write the matrix element of the exclusive transition 
$B_s\to \phi \ell^+ \ell^-$ as  
\bea
{\mathcal M} & = & 
\frac{G_F}{\sqrt{2}}\cdot\frac{ \alpha\lambda_t}{\pi} \cdot
\Big\{
C_9^{\rm eff}\,<\phi\,|\,\bar{s}\,\gamma^\mu\, P_L\, b\,|\,B_s> 
\left( \bar \ell \gamma_\mu \ell \right)
\nn
&-& \frac{2\bar m_b}{q^2}\,C_7^{\rm eff}\, 
<\phi\,|\,  \bar{s}\,i\sigma^{\mu \nu} q_\nu \,P_R\, b\, |\,B_s>  
\left( \bar \ell \gamma_\mu \ell \right)
\nn
&+& C_{10}\, <\phi\,|\,\bar{s}\,\gamma^\mu P_L \, b\,|\,B_s>
\left(\bar \ell \gamma_\mu \gamma_5 \ell\right)
\Big\},
\label{eq:matrix-elem}
\ena
where $C_7^{\rm eff}= C_7 -C_5/3 -C_6$.
One has to note that matrix element in Eq.(\ref{eq:matrix-elem}) contains
both a free quark decay amplitude coming from the operators
$\ord_7$, $\ord_9$ and $\ord_{10}$ (gluon magnetic penquin  $\ord_{8}$
does not contribute) and, in addition, certain long-distance effects 
from the matrix elements of four-quark operators $\ord_i\,\,(i=1,\ldots,6)$
which usually are absorbed into a redefinition of the short-distance 
Wilson-coefficients.
The Wilson coefficient  $ C_9^{\rm eff}$ effectively takes into account, first, 
the contributions  from the four-quark operators $\ord_i$ ($i=1,...,6$) and, 
second,  the nonperturbative effects coming from the $c\bar c$-resonance 
contributions which are as usual parametrized by the Breit-Wigner ansatz 
\cite{Ali:1991is}:
\bea
C_9^{\rm eff} & = & C_9 + 
C_0 \left\{
h(\hat m_c,  s)+ \frac{3 \pi}{\alpha^2}\,  \kappa\,
         \sum\limits_{V_i = \psi(1s),\psi(2s)}
      \frac{\Gamma(V_i \rightarrow l^+ l^-)\, m_{V_i}}
{  {m_{V_i}}^2 - q^2  - i m_{V_i} \Gamma_{V_i}}
\right\} 
\nn
&-& \frac{1}{2} h(1,  s) \left( 4 C_3 + 4 C_4 +3 C_5 + C_6\right)  
\nn
&-& \frac{1}{2} h(0,  s) \left( C_3 + 3 C_4 \right) +
\frac{2}{9} \left( 3 C_3 + C_4 + 3 C_5 + C_6 \right)\,,
\label{eq:C9eff}
\ena
where $C_0\equiv 3 C_1 + C_2 + 3 C_3 + C_4+ 3 C_5 + C_6$.
Here  the charm-loop function is written as
\bea 
h(\hat m_c,  s) & = & - \frac{8}{9}\ln\frac{\bar m_b}{\mu} 
- \frac{8}{9}\ln\hat m_c +
\frac{8}{27} + \frac{4}{9} x 
\nn
& - & \frac{2}{9} (2+x) |1-x|^{1/2} \left\{
\begin{array}{ll}
\left( \ln\left| \frac{\sqrt{1-x} + 1}{\sqrt{1-x} - 1}\right| - i\pi 
\right), &
\mbox{for } x \equiv \frac{4 \hat m_c^2}{ s} < 1, \nonumber \\
 & \\
2 \arctan \frac{1}{\sqrt{x-1}}, & \mbox{for } x \equiv \frac
{4 \hat m_c^2}{ s} > 1,
\end{array}
\right. 
\nn
h(0,  s) & = & \frac{8}{27} -\frac{8}{9} \ln\frac{\bar m_b}{\mu} - 
\frac{4}{9} \ln s + \frac{4}{9} i\pi,
\nonumber 
\ena
where $\hat m_c=\bar m_c/m_1$, $s=q^2/m_1^2$ and $\kappa=1/C_0$.
In what follows we drop the charm resonance contributions
by putting  $\kappa=0$. We will use the value of $\mu=\bar m_{b\,\rm pole}$ for 
the renormalization scale. 
Besides the charm-loop perturbative contribution, two loop contributions have
been calculated in \cite{Asatryan:2001zw,Greub:2008cy}.
They effectively modify the Wilson coefficients as
\bea
 C_7^{\rm eff} &\to&  C_7^{\rm eff}
             - \frac{\alpha_S}{4\pi}\Big( C_1 F_1^{(7)} + C_2 F_2^{(7)} \Big)\,,
\nn
 C_9^{\rm eff} &\to&  C_9^{\rm eff}
             - \frac{\alpha_S}{4\pi}\Big( C_1 F_1^{(9)} + C_2 F_2^{(9)} \Big)\,
\label{eq:Greub}
\ena
where the two-loop form factors $F_{1,2}^{(7,9)}$ are available
in Ref.~\cite{Greub:2008cy} as the Mathematica files.

The SM Wilson coefficients are taken from Ref.~\cite{Descotes-Genon:2013vna}.
They were computed at the matching scale $\mu_0=2 M_W$ and run down to 
the hadronic scale $\mu_b= 4.8$~GeV.
The evolution of couplings and current quark masses proceeds analogously. 
The values of the model independent input parameters
and the Wilson coefficients are listed in Table~\ref{tab:input}.
\bgroup 
\def\arraystretch{1.3}
\begin{table}[htbp]
\caption{Values of the input parameters.}
\vspace*{2mm}  
\centering
 \begin{tabular}{ccccccccc}
\hline\hline
  $m_W$ &  $\sin^2\theta_W $ &  $\alpha(M_Z)$ & 
$\bar m_c$ &  $\bar m_b$  &  $\bar m_t$ & $ \lambda_t$  & &\\
\hline
 $80.41$~GeV & $0.2313$ & $1/128.94$ & $1.27$~GeV & $4.68$~GeV & $173.3$~GeV &  
 0.041 & & \\
\hline\hline
 $C_1$ &  $C_2$ &  $C_3$ &  $C_4$ & $C_5$ &  $C_6$ &  $C^{\rm eff}_7$ &   
$C_9$ &  $C_{10}$ \\
\hline
 $-0.2632$ & $1.0111$ &  $-0.0055$ & $-0.0806$ & 0.0004 & 0.0009 & $-0.2923$ &
  4.0749 & $-4.3085$ \\
\hline\hline
 \end{tabular}
\label{tab:input}
\end{table}
\egroup

A global analysis
of $b\to s\ell\ell$ anomalies has been performed in 
Ref.~\cite{Descotes-Genon:2015uva} with the Next-to-Next-to-Leading Logarithmic (NNLL) corrections included.
It was shown that they amount up to 15$\%$. The discussion of 
the non-local $c\bar c$ contributions maybe also found in
Ref.~\cite{Beylich:2011aq}.

\section{Numerical results}

We are aiming to compare our results for the branching fractions and angular
observables with the experimental data recently reported by
the LHCb Collaboration  \cite{Aaij:2015esa} and the results of global analyses
performed in Ref.~\cite{Descotes-Genon:2015uva}. 
The four-fold distribution in the cascade decay 
$B\to\phi(\to K^+K^-)\bar \ell \ell$ allows one to define a number
of physical observables which can be measured experimentally.
The observables accessible in the decay $B_s\to\phi\mu^+\mu^-$  
\cite{Aaij:2015esa} are the CP~averaged differential branching ratio
$d{\cal B}/dq^2$, the CP-averaged $\phi$ longitudinal polarization
fraction $F_L$, forward-backward asymmetry $A_{FB}$
and the CP-averaged angular observables $S_{3,4,7}$
which may be related to the optimized observables $P_i$ 
\cite{Descotes-Genon:2015uva}. 
The CP~asymmetries $A_{5,6,8,9}$  \cite{Bobeth:2008ij} in the SM  
are induced by the weak phase from the CKM matrix. 
For the $b\to s$ transitions the CP~asymmetries are proportional to 
${\rm Im}(\hat\lambda_u)\equiv {\rm Im}(V_{ub}V^\ast_{us}/V_{tb}V^\ast_{ts})$ 
which is of order $10^{-2}$~\cite{Bobeth:2008ij}. The experimental data 
reported by  \cite{Aaij:2015esa}
contain huge statistical uncertainties (see, Table 3 in \cite{Aaij:2015esa}). 
For these reasons we restricted ourselves to the CP-averaged quantities. 

We start with the branching fraction of the rare decay 
$B_s\to \phi\bar \ell \ell$. The width of this decay is computed 
 by integration of the  $q^2$-differential distribution
\bea
&&
\frac{d\Gamma(B\to \phi\bar \ell \ell)}{dq^2} =\,
 \frac{G^2_F}{(2\pi)^3}\,
\left(\frac{\alpha \lambda_t}{2\pi}\right)^2
\frac{|{\bf p_2}|\,q^2\,\beta_\ell}{12\,m_1^2}
{\cal H}_{\rm tot}\,,
\nn[1.2ex]
&&
{\cal H}_{\rm tot} = 
\frac12\left(  {\cal H}^{11}_U + {\cal H}^{22}_U 
             + {\cal H}^{11}_L + {\cal H}^{22}_L \right )
+ \delta_{\ell\ell}\,
\left[\,\frac12 {\cal H}^{11}_U - {\cal H}^{22}_U
       + \frac12 {\cal H}^{11}_L - {\cal H}^{22}_L + \frac32\, {\cal H}^{22}_S
\right].
\label{eq:distr1}
\ena
In what follows we will use the short notation
$m_1=m_{B_s}$, $m_2=m_\phi$, 
$\beta_\ell=\sqrt{1-4m_\ell^2/q^2}$, $\delta_{\ell\ell} = 2m^2_\ell/q^2$.
Then  $|{\bf p_2}|=\lambda^{1/2}(m_1^2,m_2^2,q^2)/(2\,m_1)$ is the momentum of 
the $\phi$-meson given in the $B_s$-rest frame.
The bilinear combinations of the helicity amplitudes ${\cal H}$ are defined 
as (see, Ref.~\cite{Faessler:2002ut} for details): 
\bea
{\cal H}^{ii}_U   &=&  |H^i_{+1 +1}|^2 +  |H^i_{-1 -1}|^2, \qquad
{\cal H}^{ii}_L   = |H^i_{00}|^2, \qquad 
{\cal H}^{ii}_S   = |H^i_{t0}|^2 ,
\label{eq:bilinear}
\ena
where the helicity amplitudes are expressed via the form factors 
appearing in the matrix element of the rare decay
$B_s\to \phi\bar \ell \ell$ as  
\bea
H^i_{t0} &=& 
\frac{1}{m_1+m_2}\frac{m_1\,|{\bf p_2}|}{m_2\sqrt{q^2}}
         \left(Pq\,(-A^i_0+A^i_+)+q^2 A^i_-\right),
\nn[1.2ex]
H^i_{\pm1\pm1} &=& 
\frac{1}{m_1+m_2}\left(-Pq\, A^i_0\pm 2\,m_1\,|{\bf p_2}|\, V^i \right),
\nn[1.2ex]
H^i_{00} &=&  
\frac{1}{m_1+m_2}\frac{1}{2\,m_2\sqrt{q^2}} 
\left(-Pq\,(m_1^2 - m_2^2 - q^2)\, A^i_0 + 4\,m_1^2\,|{\bf p_2}|^2\, A^i_+\right).
\label{eq:hel_V}
\ena
The form factors $A^i$ and $V^i$ $(i=1,2)$ are related to the
form factors in the  $B_s-\phi$ transitions, see Eqs.~(\ref{eq:PV})
and (\ref{eq:PVT}), in the following manner
\bea 
V^{(1)} &=&   C_9^{\rm eff}\,V  + C_7^{\rm eff}\,g \,\frac{2\bar m_b(m_1+m_2)}{q^2}\,,
\nn
A_0^{(1)} &=& C_9^{\rm eff}\,A_0 
+ C_7^{\rm eff}\,a_0\,\frac{2\bar m_b(m_1+m_2)}{q^2}\,,
\nn
A_+^{(1)} &=& C_9^{\rm eff}\,A_+ 
+ C_7^{\rm eff}\,a_+\,\frac{2\bar m_b(m_1+m_2)}{q^2}\,,
\nn
A_-^{(1)} &=& C_9^{\rm eff}\,A_- 
+ C_7^{\rm eff}\,(a_0-a_+)\,\frac{2\bar m_b(m_1+m_2)}{q^2}\,\frac{Pq}{q^2}\,,
\nn[1.5ex]
V^{(2)}   &=& C_{10}\,V, \qquad A_0^{(2)} = C_{10}\,A_0,\qquad
A_\pm^{(2)} = C_{10}\,A_\pm.
\label{eq:ff-relations}
\ena

The differential rate of the decay $B_s\to\phi\nu\bar\nu $ is calculated 
according to

\be
\frac{d\Gamma(B_s\to\phi\nu\bar\nu)}{dq^2} 
= \frac{G_F^2}{(2\pi)^3} \Big(\frac{\alpha\lambda_t}{2\pi}\Big)^2
\Big[\frac{D_\nu(x_t)}{\sin^2\theta_W}\Big]^2
\frac{|{\bf p_2}|\, q^2}{4m_1^2}\cdot (H_U+H_L)\,,
\en
where $x_t=\bar m_t^2/m_W^2$ and the function $D_\nu$ is given by
\be
D_\nu(x) = \frac{x}{8}\left(\frac{2+x}{x-1}+\frac{3x-6}{(x-1)^2}\,\ln x\right).
\en
The relevant bilinear helicity combinations are defined as
\bea
{\cal H}_U   &=&  |H_{+1 +1}|^2 +  |H_{-1 -1}|^2, \qquad 
{\cal H}_L   = |H_{00}|^2, 
\nn[1.2ex]
H_{\pm1\pm1} &=& 
\frac{1}{m_1+m_2}\left(-Pq\, A_0\pm 2\,m_1\,|{\bf p_2}|\, V \right),
\nn[1.2ex]
H_{00} &=&  
\frac{1}{m_1+m_2}\frac{1}{2\,m_2\sqrt{q^2}} 
\left(-Pq\,(m_1^2 - m_2^2 - q^2)\, A_0 + 4\,m_1^2\,|{\bf p_2}|^2\, A_+\right).
\label{eq:bilinear-2}
\ena

The width of the color-suppressed nonleptonic decay
$B_s\to\Jpsi\,\phi$ decay is given by \cite{Ivanov:2011aa} 
\bea
\Gamma(B_s\to\Jpsi\,\phi ) &=&
\frac{G_F^2}{16\pi}\frac{|{\bf p_{\,2}}|}{m^2_{1}}  
|V_{cb}V_{cs}|^2 
\left(C^{\,\rm eff}_1+ C^{\,\rm eff}_5\right)^2 
\left( m_{\Jpsi}\,f_{\Jpsi} \right)^2\,(H_U+H_L)
\label{eq:BsJpsiPhi}
\ena
where the momentum transfer squared is taken on the mass of $\Jpsi$, i.e.
$q^2=m^2_{\Jpsi}$, $V_{cb}~=~0.406$, $V_{cs}=0.975$ and $f_{\Jpsi}=415$~MeV. The  Wilson coefficients are combined as
$ C^{\,\rm eff}_{1}=C_1+\xi\, C_2+C_3+\xi\, C_4 $ and
$ C^{\,\rm eff}_{5}=C_5+\xi\, C_6$ in accordance with the naive factorization.
The terms multiplied by the color factor $\xi=1/N_c$  will be dropped 
in the numerical calculations according to the $1/N_c$-expansion.

Finally, we calculate the width of radiative decay $B_s\to\phi\gamma $ defined 
by
\be
\Gamma(B_s\to\phi\gamma)
= \frac{G_F^2\alpha\lambda_t^2}{32\pi^4}
\bar m_b^2 m_1^3\Big(1-\frac{m_2^2}{m_1^2}\Big)^3\,|C^{\rm eff}_7|^2\, g^2(0)\,.
\en

One has to note that the experimental observables in the decays
of neutral $B_s$-mesons are affected by $B_s-\bar B_s$ mixing.
The theoretical framework for studying the time-dependent decays
with taking into account such mixing has been recently developed 
in Ref.~\cite{Descotes-Genon:2015hea}. The mixing effects
change the values of rates and CP averages within a few percents.

In Table~\ref{tab:branching} the calculated values of branching fractions
$B_s\to \phi\mu^+\mu^- $,  $B_s\to  \phi\tau^+\tau^- $, $B_s\to  \phi\gamma $,
$B_s\to  \phi\nu\bar\nu $ and $B_s\to  \phi J/ \psi$ are given.
The experimental errors shown in  Table~\ref{tab:branching} result from 
combining the partial uncertainties in quadrature.
The model uncertainties are estimated to be within 10$\%$. 
We compare our results with those obtained in other approaches.

\begin{table}[ht]
\caption{Total branching fractions. }
\label{tab:branching}
\begin{ruledtabular}
\begin{tabular}{l|llllll}
 & This work & Ref.~\cite{Faustov:2013pca} & Ref.~\cite{Yilmaz:2008pa} 
 &  Ref.~\cite{Wu:2006rd} & Ref.~\cite{Geng:2003su} 
 & Ref.~\cite{Agashe:2014kda,Aaij:2015esa} \\
\hline
 $10^7 {\cal B}(B_s\to\phi \mu^+\mu^-)$ & $9.11\pm 1.82$ & $11.1\pm 1.1$
 &  19.2  & $11.8\pm 1.1$ & 16.4 &   $ 7.97 \pm 0.77 $ \\
\hline 
$10^7 {\cal B}(B_s\to\phi \tau^+\tau^-)$ & $1.03\pm 0.20$ & $1.5\pm 0.2$ 
 & 2.34 & $1.23\pm 0.11$ & 1.51  & \\
\hline  
$10^5 {\cal B}(B_s\to\phi\gamma) $ & $ 2.39\pm 0.48$ & $3.8\pm 0.4$ 
 & & & & $3.52\pm 0.34$ \\
\hline 
 $10^5 {\cal B}(B_s\to\phi \nu\bar\nu) $   & $0.84\pm 0.16$ 
 &  $0.796\pm 0.080 $& && 1.165 & $< 540 $ \\
\hline 
 $10^2 {\cal B}(B_s\to\phi J/\psi)$ & $0.16\pm 0.03$ &  $0.113\pm 0.016$
 &&&& $0.108\pm 0.009$ \\
\end{tabular}
\end{ruledtabular}
\end{table}

The full four-fold angular decay distribution 
for the rare $B$~decay has been derived in Ref.~ \cite{Faessler:2002ut} 
in terms of helicity amplitudes including lepton mass effects.
It is described by the three angles and the squared momentum $q^2$
of the lepton pair. This distribution allows one to define a number of physical
observables which can be measured experimentally.
Among them are three natural observables: the branching ratio, 
the longitudinal polarization fraction of the $\phi$-meson and 
the forward-backward asymmetry.
The differential branching fraction is obtained from
the full four-fold angular decay distribution by integration over
all three angles. The explicit expression is given by Eq.~(\ref{eq:distr1})
in terms of helicity amplitudes. The relation of helicity amplitudes
and the transversality amplitudes is obtained in Ref.~\cite{Dubnicka:2015iwg}.
 
The  longitudinal polarization fraction and the forward-backward asymmetry
are defined as
\bea
F_L &=&  
\frac12 \beta_\ell^2 
\frac{ {\cal H}_L^{11} +  {\cal H}_L^{22}}{ {\cal H}_{\rm tot} }, 
\label{eq:FL}\\[2ex]
F_T &=&  
\frac12 \beta_\ell^2 
\frac{ {\cal H}_U^{11} +  {\cal H}_U^{22}}{ {\cal H}_{\rm tot} },
\label{eq:FT}\\[2ex]
A_{\rm FB} &=& 
\frac{1}{d\Gamma/dq^2} \left[ \int\limits_0^1 - \int\limits_{-1}^0 \right]
d\!\cos\theta\, \frac{d^2\Gamma}{dq^2 d\!\cos\theta} 
= -\frac34\beta_\ell \frac{ {\cal H}_P^{12}} { {\cal H}_{\rm tot} }\,,
\label{eq:AFB}
\ena
where $\theta$ is the polar angle between the $\ell^+\ell^-$-plane and z-axis.
As follows from the definition, the quantities $A_{\rm FB}$ and $F_L$
are the ratios of the hadronic amplitudes which are supposed to be less
dependent on the theoretical uncertainties.
 
\begin{figure*}[htbp]
\centering
\begin{tabular}{lr}
\includegraphics[scale=0.6]{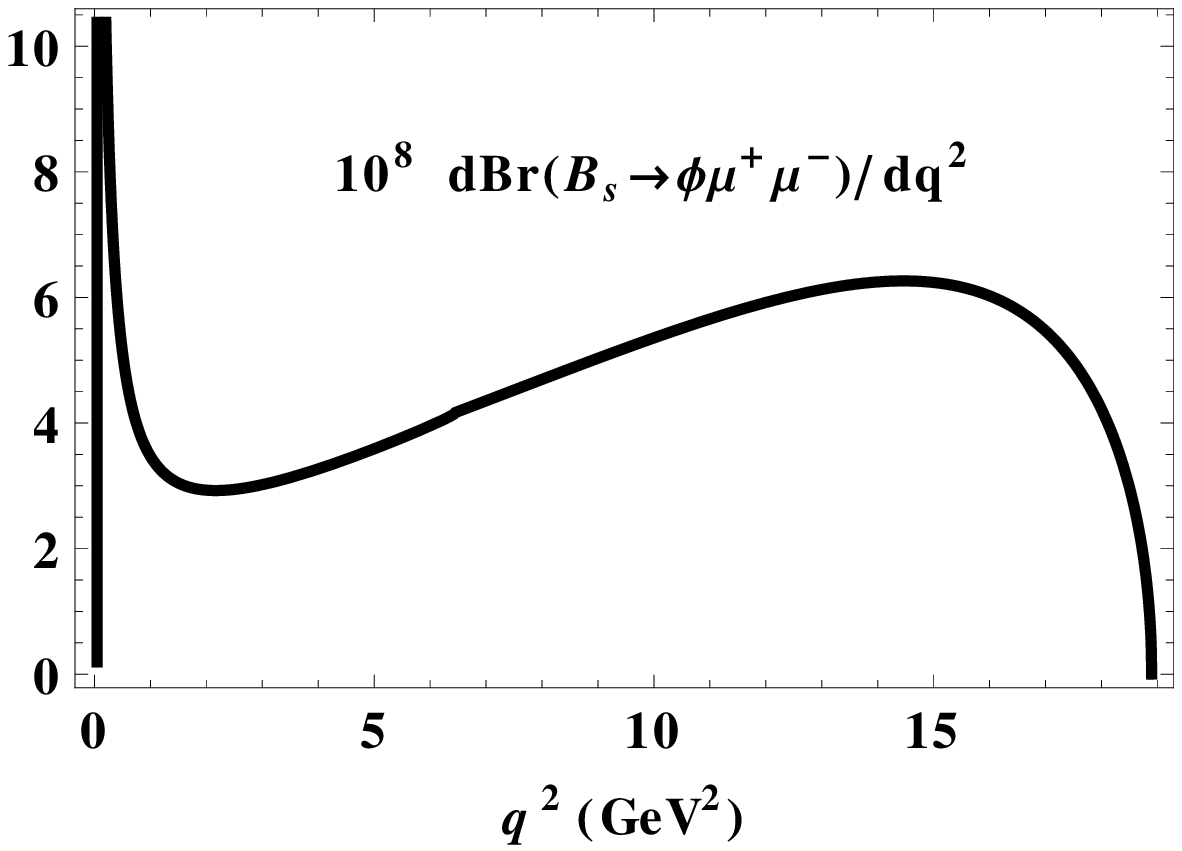} &
\includegraphics[scale=0.6]{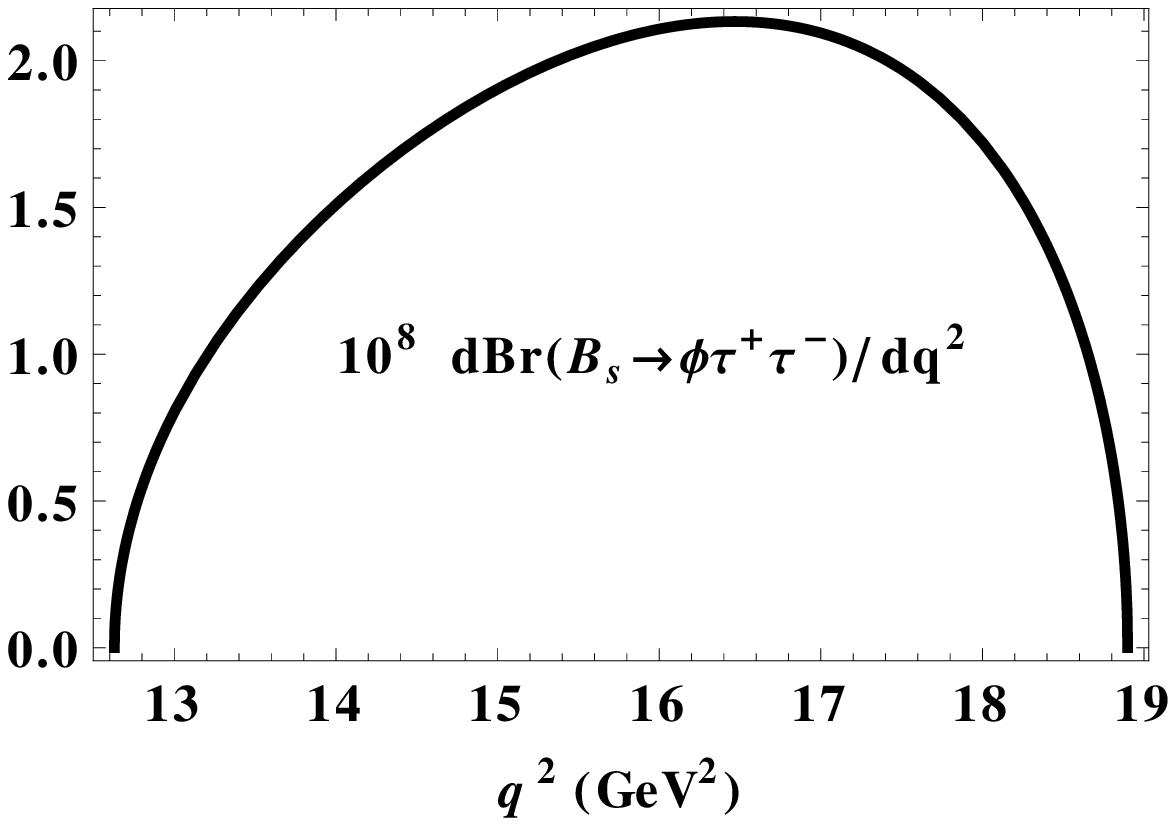}\\
\end{tabular}
\caption{Differential branching fraction in GeV$^{-2}$.}
\label{fig:Br}
\end{figure*}

\begin{figure*}[htbp]
\centering
\begin{tabular}{lr}
\includegraphics[scale=0.6]{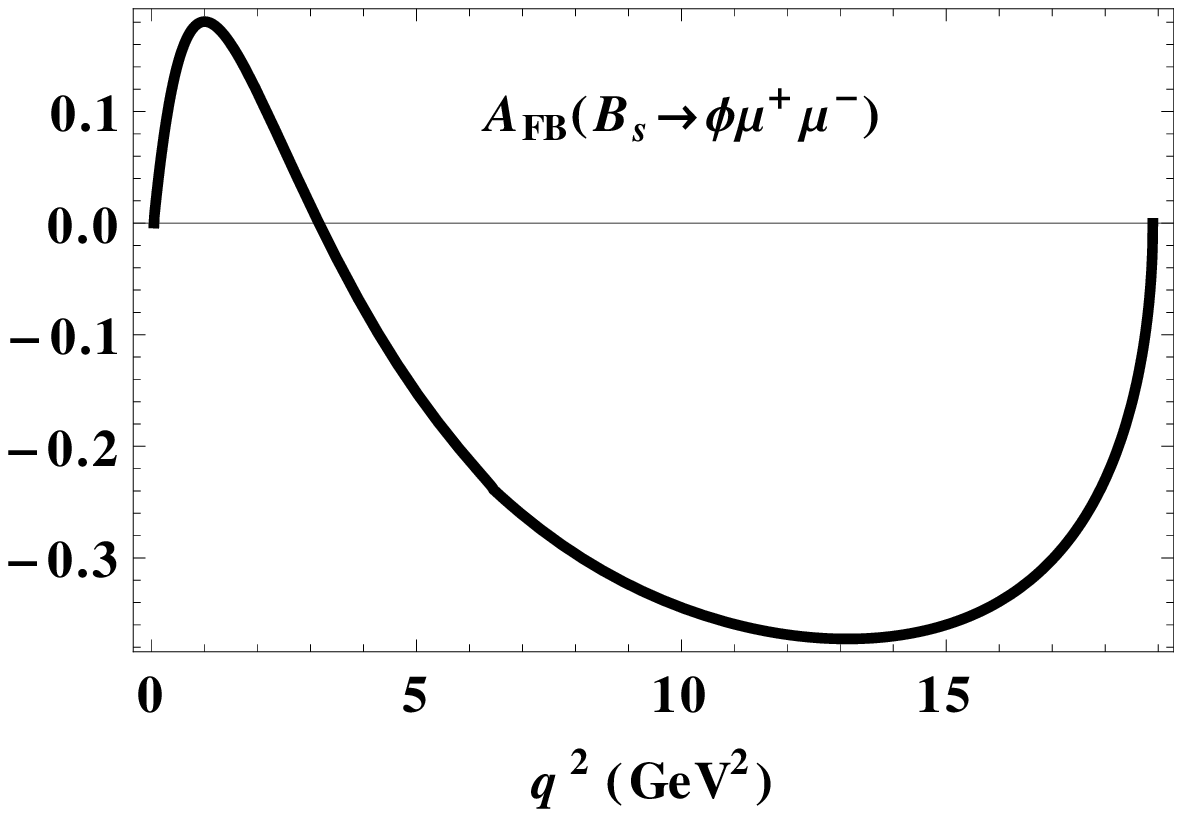} &
\includegraphics[scale=0.6]{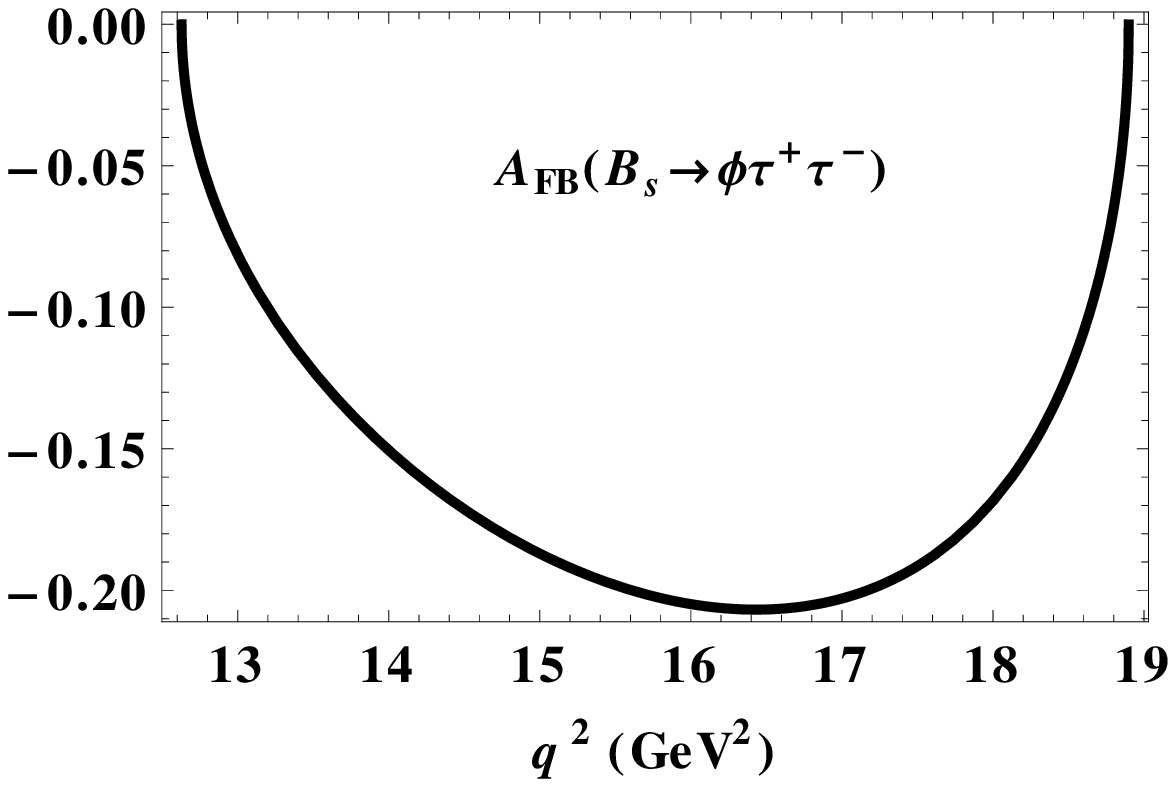}\\
\end{tabular}
\caption{Forward-backward asymmetry $A_{FB}$.}
\label{fig:AFB}
\end{figure*}

\begin{figure*}[htbp]
\centering
\begin{tabular}{lr}
\includegraphics[scale=0.6]{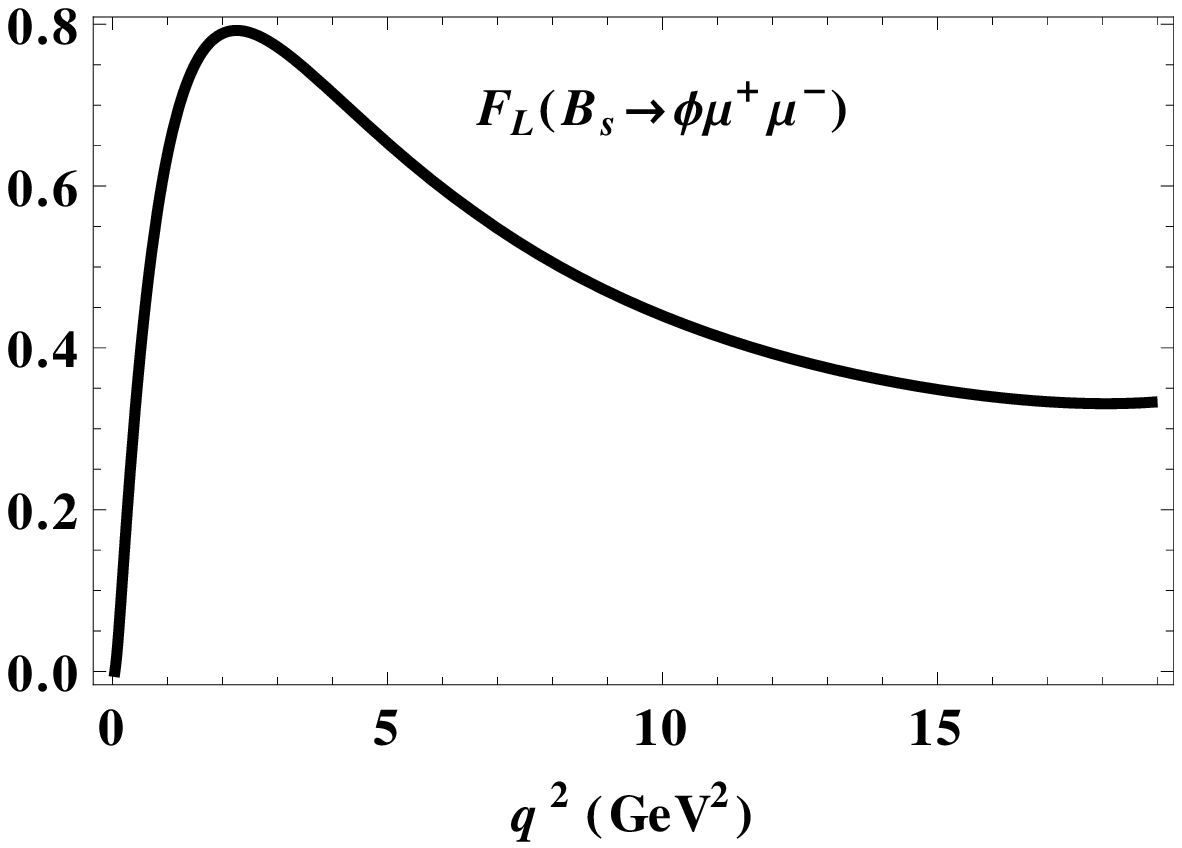} &
\includegraphics[scale=0.6]{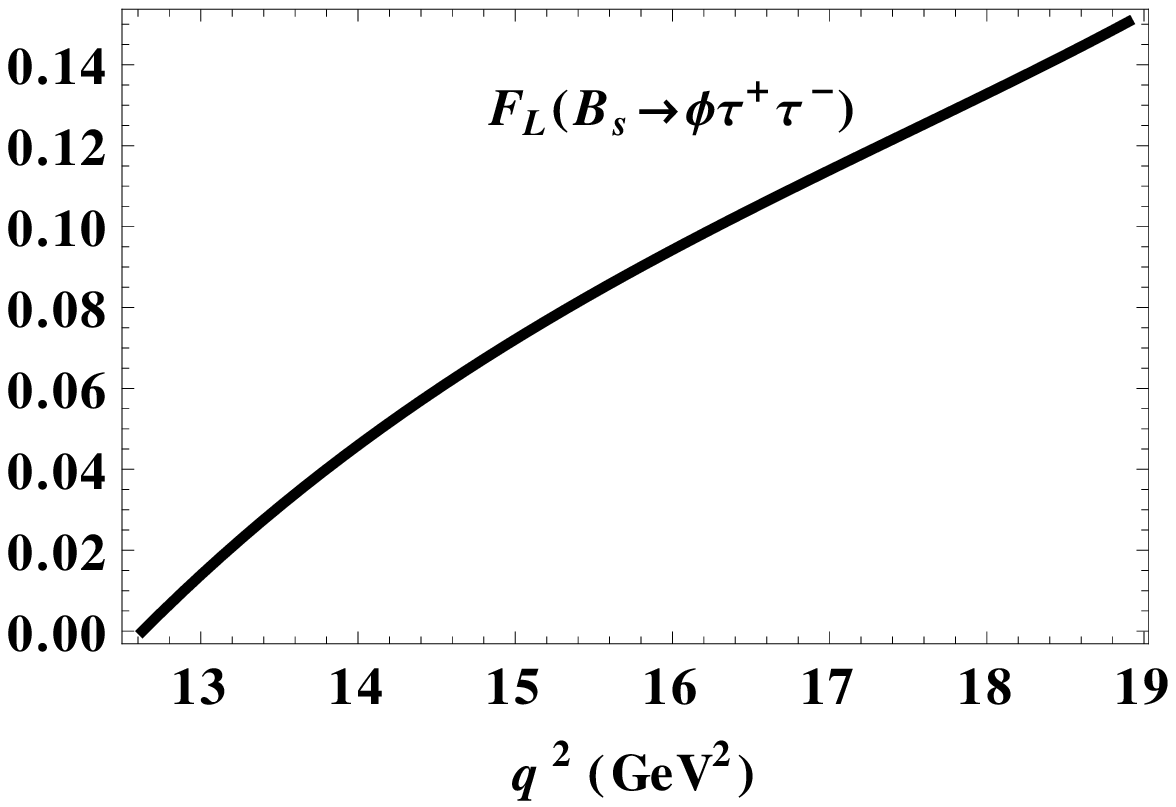}\\
\end{tabular}
\caption{Longitudinal polarization $F_L$.}
\label{fig:FL}
\end{figure*}

The behavior of the differential branching fraction $d {\cal B}/dq^2$,
forward-backward asymmetry $A_{FB}$ and  longitudinal polarization $F_L$
is shown in Figs.~\ref{fig:Br},  \ref{fig:AFB} and \ref{fig:FL},
respectively. 

A set of so-called optimized observables $P_i$ has been
constructed  (see \cite{Descotes-Genon:2013vna}
and references therein) by taking appropriate ratios
of the form factors in such a way to minimize the hadronic 
uncertainties.  These observables have been constructed with 
aim to reduce the form factor dependence and increase the discrimination power 
between the SM and NP, together with preserving a good experimental 
accessibility. It seems however more difficult to give them a clean physical 
interpretation, as it was the case for $A_{FB}$ and $F_L$.

The optimized observables  have not been given explicitly in 
\cite{Aaij:2015esa}.
Their numerical values were obtained in  \cite{Descotes-Genon:2015uva}
by converting the results for the CP~averages  $S_{3,4,7}$ 
into the optimized observables. 

\begin{figure*}[htbp]
\centering
\begin{tabular}{lr}
\includegraphics[scale=0.6]{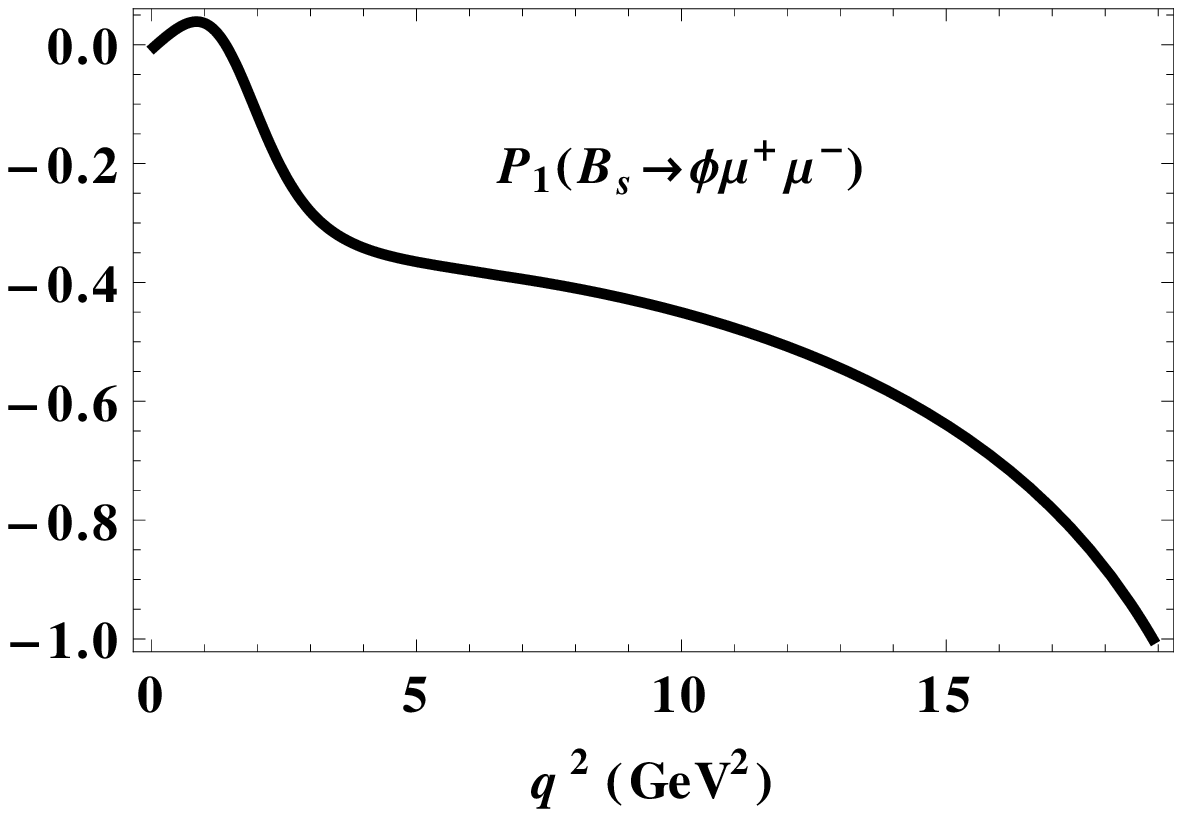} &
\includegraphics[scale=0.6]{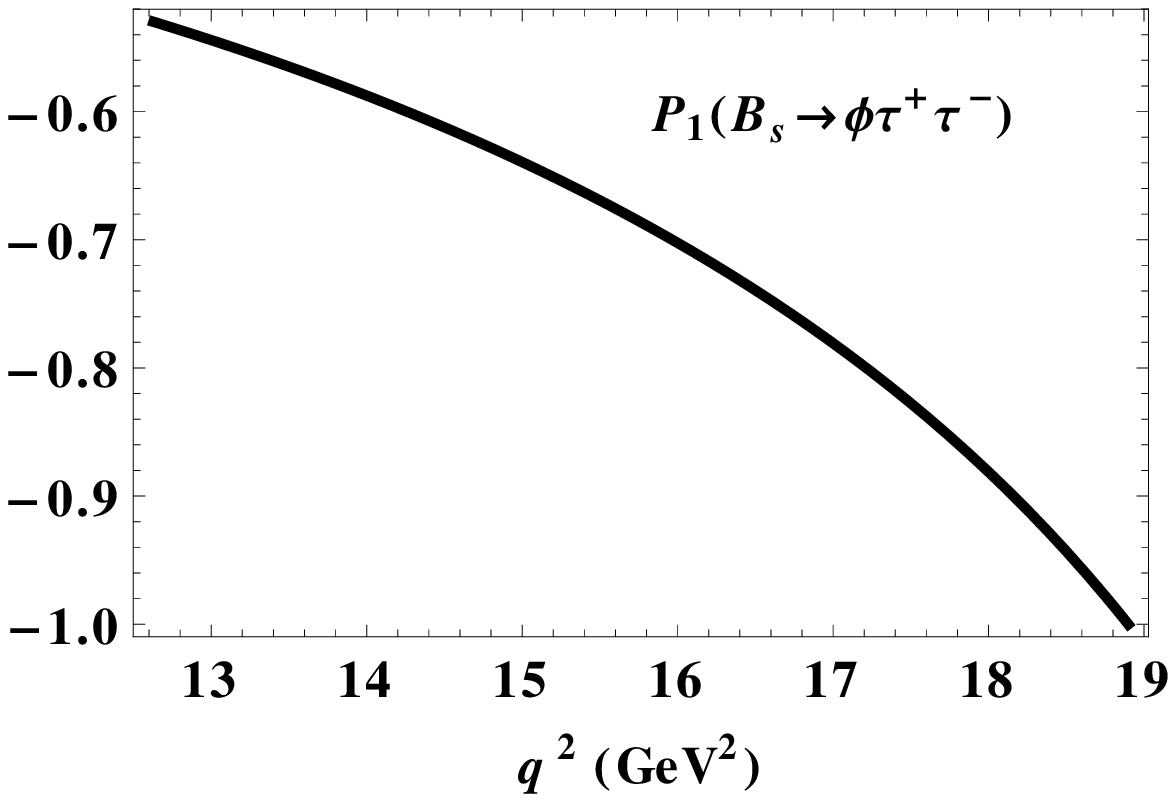}\\
\includegraphics[scale=0.6]{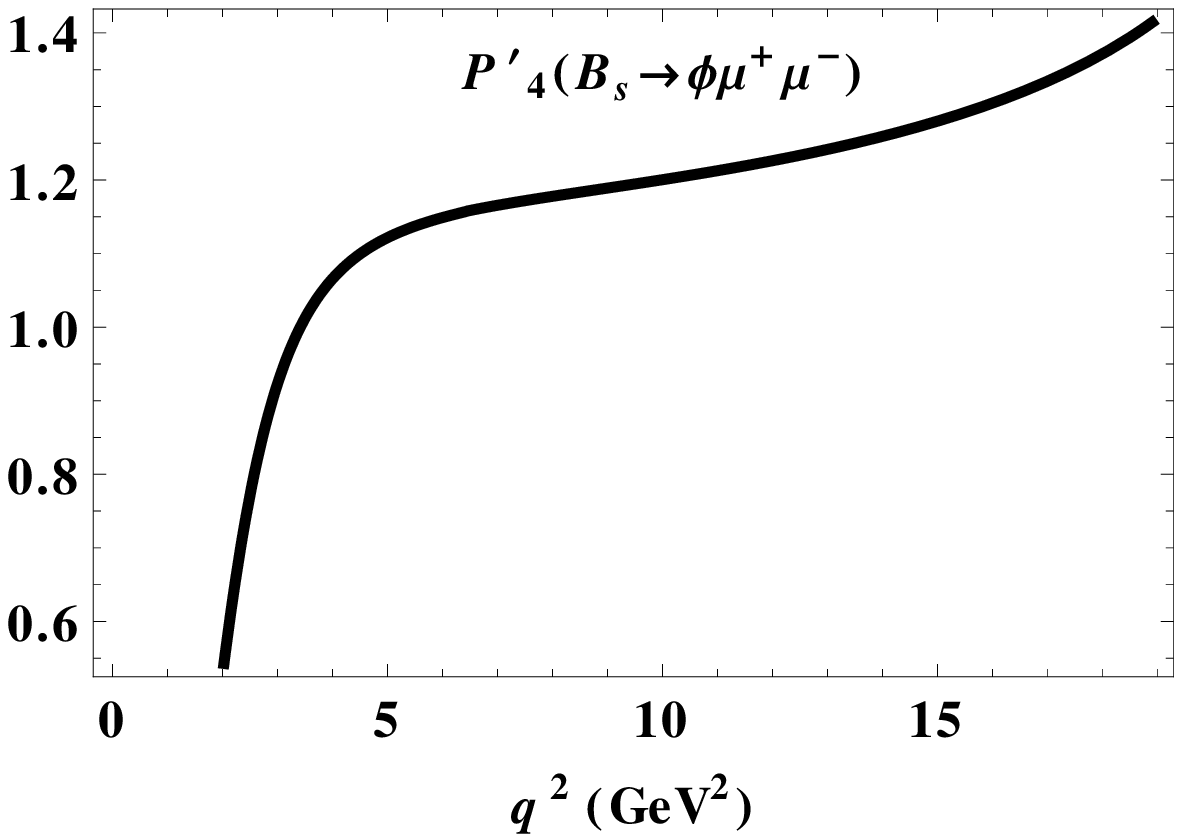} &
\includegraphics[scale=0.6]{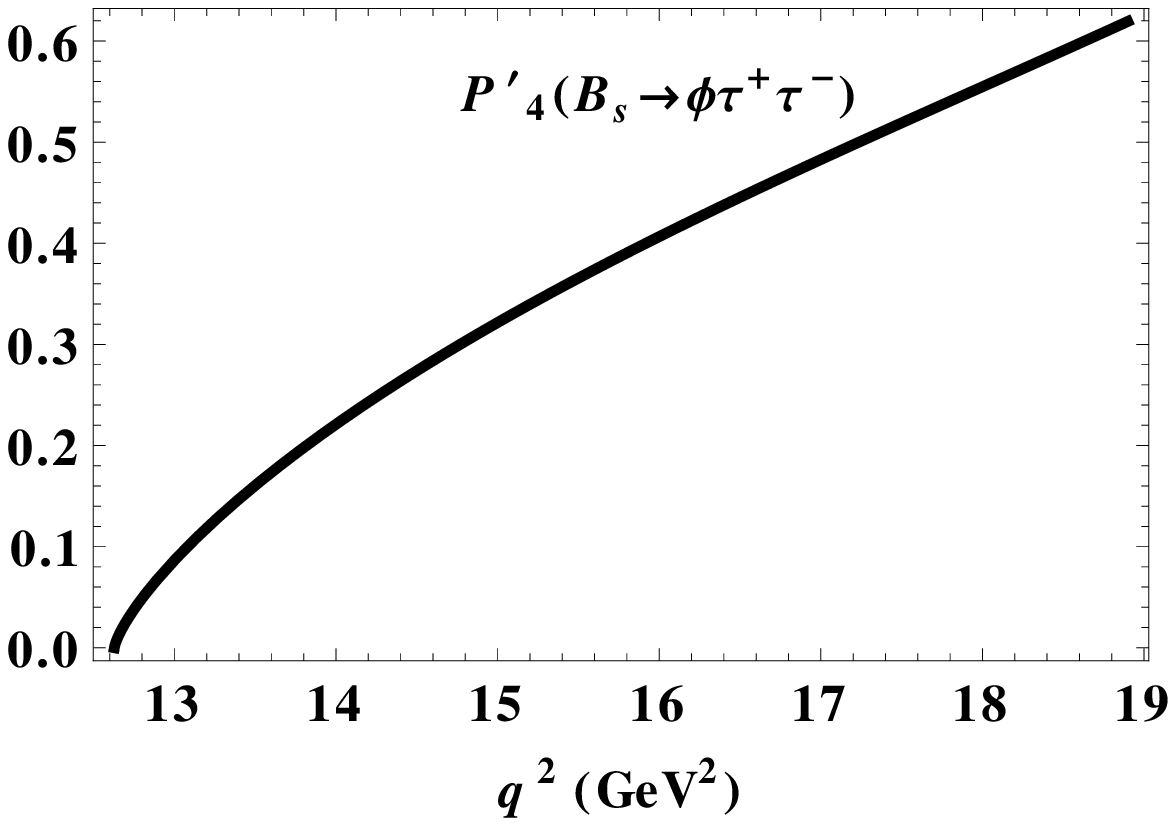}\\
\end{tabular}
\caption{Clean observables $P_{1}$ and  $P'_4$.}
\label{fig:P14}
\end{figure*}

We will calculate directly the optimized observables $P_i$ expressed
through the helicity amplitudes as was done in Ref.~\cite{Dubnicka:2015iwg}.
The $q^2$-dependence  of the optimized observables $P_1$ and $P'_4$ 
is displayed  in Fig.\ref{fig:P14}. 

The {$q^{2}$--averages of  polarization observables over 
the whole  allowed kinematic region are given in Table~\ref{tab:obs-numerics}.
For comparison reasons, we also give the values of the  $S_{3,4,7}$ by using 
the relations \cite{DescotesGenon:2012zf}:
\be
S_3 = \frac12 F_T P_1, \qquad 
S_4 = \frac12 \sqrt{F_T F_L} P'_4, \qquad
S_7 = - \sqrt{F_T F_L} P'_6 .
\en

\begin{table}[ht] 
\begin{center}
\caption{$q^{2}$--averages of polarization observables over the whole 
allowed kinematic region. }
\label{tab:obs-numerics}
\vspace{2mm}
\def\arraystretch{1.2}
\begin{tabular}{c|cccccc}
\hline
\multicolumn{7}{c}{ $B_s\to\phi\ell^+\ell^-$ }\\
\hline
& $<A_{FB}>$       
& $<F_L>$ 
& $<P_1>$   
& $<P_4'>$ 
& $<S_3>$ 
& $<S_4>$
\\
\hline
 $\mu$ \qquad   
& $ -0.24\pm 0.05 $  
& $  0.45\pm 0.09 $  
& $ -0.52\pm 0.1$   
& $  1.05\pm 0.21$   
&$ -0.14\pm 0.03$
& $ 0.26\pm 0.05$   
\\
 $\tau$ \qquad    
& $ -0.18\pm 0.04$   
& $  0.090\pm 0.02$  
& $ -0.76\pm 0.15$    
& $ 1.33\pm 0.27$   
& $ -0.067\pm 0.013$
&  $ 0.083\pm 0.017$      
\\
\hline
\end{tabular}
\end{center}
\end{table}

We use the Wilson coefficients obtained at the next-to-leading logarithmic
(NLL) order in our calculation of the observables in the full kinematical
region of the momentum transfer squared. At this order only 
the coefficient $C_9^{\rm eff}$ has imaginary part. Since our form factors are 
real, the optimized observable $P'_6$ is identically zero at this order. 
The optimized  observable $P_1$ is small for large recoil for any
choice of the Wilson coefficients. It is easy to check that
$P_1\propto A_0(0)-V(0)$ at $q^2=0$. In our model $A_0(0)=0.40$ and $V(0)=0.31$
so it leads to the really small value of the $P_1$. Note that 
$A_0(0)=V(0)$ in the heavy quark limit.

Finally, we present our results for the binned observables in
Table~\ref{tab:bin}. Here, we take into account the NNLL corrections
for the Wilson coefficients which have been calculated in 
\cite{Asatryan:2001zw,Greub:2008cy}.
They effectively modify the Wilson coefficients $C_7^{\rm eff}$ and 
$C_9^{\rm eff}$. 
The two-loop corrections to the decay $b\to s\ell^+\ell^-$
were given in \cite{Asatryan:2001zw} as expansions in
the small parameters $\hat s=q^2/\bar m_b^2$ and $z=\bar m_c^2/\bar m_b^2$.
The $q^2$-region was restricted to the range $0.05\le \hat s \le 0.25$.
The NNLL corrections in the high $q^2$ region above the charm threshold
$q^2>4\bar m_c^2$ were presented in \cite{Greub:2008cy}. 
The high $q^2$-region was restricted to the range $0.4 \le \hat s \le 1.0$.
By using the value of QCD bottom quark $\bar m_b=4.68$~GeV from 
Table~\ref{tab:input} one can obtain the  $q^2$ regions where the two-loop 
corrections are valid:
\be
1.1 \le q^2 \le 5.5\,\,\text{GeV}^2 \quad\text{(low region)}
\quad
\text{and}
\quad
8.8 \le q^2 \le 22\,\,\text{GeV}^2 \quad\text{(high region)}.
\label{eq:NNLL}
\en
However, for the 2-loop calculation in the low $q^2$ region, it was important
that $q^2/{\bar m_b}^2$ and $q^2/(4 \bar m_c^2)$ are both much smaller than 1.
So one can safely use the results of Ref.~\cite{Asatryan:2001zw} 
for low $q^2$  in the region $0.1\le q^2\le 6$~GeV$^2$.
The two-loop expansion in the high $q^2$ region  \cite{Greub:2008cy}
can be justified only for $q^2/{\bar m_b}^2>0.4$.
For this reason, we exclude the bin $[5,8]$ from the consideration
of two-loop corrections
\footnote{We appreciate Christoph Greub for discussion of these points.}.

Now the observable $P'_6$ and hence $S_7$ become different from zero. 
The NNLL corrections contribute up to
20$\%$ in the region of small transferred momentum squared $q^2\le 6$~GeV$^2$ 
but their influence in the region of large  $q^2$ is really negligible.

\section{Discussion and conclusion}

The level of agreement with experiment can be estimated by combining in 
quadrature the experimental errors with the theoretical ones: if the 
difference in observable values is smaller, then it can be seen as 
compatible with zero.

Using this optics one can address the $3.3~\sigma$ deviation seen by 
\cite{Aaij:2015esa} for branching fraction in the $1-6~\mathrm{GeV^2}$ range. 
In the covariant confined quark model this discrepancy is much reduced. 
The remaining deviation ($  0.9~\sigma$) shrinks even further if 
the two-loop corrections for the Wilson coefficients are taken into account, 
down to $  0.7~\sigma$. With such error reduction one cannot claim a 
discrepancy with the SM any longer.

Overall one observes a good description of the data by the covariant quark 
model and the agreement becomes even better if the two-loop corrections are 
taken into account. The biggest discrepancy of $  1.9~\sigma$ observed for 
$F_L$ in the lowest bin $0.1 \le q^2 \le 2 ~\mathrm{GeV^2}$ is reduced to $1.4~\sigma$ 
when these corrections are taken into account. 

The remaining deviations do not exceed $1.7~\sigma$ and only one of them is 
greater than $ 1.4~\sigma$ if the two loops corrections are neglected 
( $S_4$  for $15 \le q^2 \le 17 ~\mathrm{GeV^2}$). When they 
are taken into account most measurements lie within one standard deviation, 
the only one exceeding $ 1.4~\sigma$ is $S_4$ for $15 \le q^2 \le 17 ~\mathrm{GeV^2}$.

The largest deviation of the obtained results from 
the SM~\cite{Descotes-Genon:2015uva} predictions is
found for the branching fraction at lower $q^2$. 
One has to emphasize that the branching fraction is the most
affected by the uncertainties related to the hadronic form factors. 
The global analysis performed in \cite{Descotes-Genon:2015uva}
has basically used a specific set of form factors determined
from light-cone sum rules (LCSR)  
\cite{Straub:2015ica,Khodjamirian:2006st,Ball:2004rg}.

As discussed above, the value of $P_1$ at small $q^2$
is really small and lays within uncertainties given by both the experiment
and global fit \cite{Descotes-Genon:2015uva}. There is agreement
between our approach and  \cite{Descotes-Genon:2015uva} for large $q^2$.
The values of $P'_6$ are identical zero at one-loop level.
The results obtained by using two-loop expansion are in agreement
with the experiment  and with \cite{Descotes-Genon:2015uva} within uncertainties.

One can conclude that the results provided by the covariant confined quark 
model do not allow to claim a significant deviation from the SM  and they 
demonstrate the non-negligible effect of the two-loop corrections for the 
Wilson coefficients which bring the theoretical predictions closer to the data.

\clearpage

\begin{longtable}{@{}ccccc@{}}
\caption{Binned observables.}\\
\label{tab:bin}
\endfirsthead
\endhead
\toprule[1.6pt] 
$ 10^7 {\cal B}(B_s\to \phi\mu^+\mu^-) $  & 2 loop & 1 loop 
& SM~\cite{Descotes-Genon:2015uva} & Expt.~\cite{Aaij:2015esa}\\ 
 \midrule 
 $[0.1,2]$ & $0.99\pm 0.2$   & $0.86\pm 0.17 $ 
         & $1.81 \pm 0.36$ & $1.11 \pm 0.16$  \\ 
 $[2,5]$   & $0.90\pm 0.18$  & $0.95\pm 0.19 $  
         & $1.88\pm 0.31$  & $0.77\pm 0.14$   \\ 
 $[5,8]$   &  $--$           & $1.25\pm 0.25$
         &  $2.25\pm 0.41$ & $0.96\pm 0.15$   \\ 
 $ [11,12.5]$ & $0.84\pm 0.17$&$0.88\pm 0.18$
  				& $--$ & $ 0.71 \pm 0.12 $  \\ 
 $ [15,17]$ & $1.15\pm 0.23$&$1.19\pm 0.24$
 				& $--$ & $ 0.90 \pm 0.13 $  \\ 
 $ [17,19]$ & $0.75\pm 0.15$&$0.77\pm 0.15$
 				& $--$ & $ 0.75 \pm 0.13 $ \\ 
 $ [1.,6.]$ & $1.56\pm 0.31$&$1.64\pm 0.33$
 				& $--$ & $ 1.29 \pm 0.19 $ \\ 
 $[15,19]$ & $1.89\pm 0.28$  & $1.95\pm 0.29$ 
         & $2.20\pm 0.16$  & $1.62\pm 0.20$   \\ 
\midrule[1.6pt] 
$ F_L (B_s\to \phi\mu^+\mu^-) $  & 2 loop & 1 loop 
& SM~\cite{Descotes-Genon:2015uva} & Expt.~\cite{Aaij:2015esa} \\ 
 \midrule 
 $[0.1,2]$ & $0.37\pm 0.07$  & $0.46\pm 0.09$     
         & $0.46\pm 0.09$  & $0.20\pm 0.09$   \\ 
 $[2,5]$   & $0.72\pm 0.14$  & $0.74\pm 0.15$
         & $0.79\pm 0.03$  & $0.68\pm 0.15$   \\ 
 $[5,8]$   & $--$ & $0.57\pm 0.11$
         & $0.65\pm 0.05$  & $0.54\pm 0.10$   \\ 
  $ [11,12.5] $ & $0.40\pm 0.08$&$0.40\pm 0.08$
  				& $--$ & $ 0.29 \pm 0.11 $  \\ 
 $ [15,17] $ 	& $0.34\pm 0.07$&$0.34\pm 0.07$
 				& $--$ & $ 0.23 \pm 0.09 $ \\ 
 $ [17,19] $ 	& $0.33\pm 0.06$&$0.33\pm 0.06$
 				& $--$ & $ 0.4 \pm 0.14 $ \\ 
 $ [1,6] $ 	& $0.69\pm 0.14$&$0.71\pm 0.14$
 				& $--$ & $ 0.63 \pm 0.09 $  \\ 
 $[15,19]$ & $0.34\pm 0.07$  & $0.34\pm 0.07$ 
         & $0.36\pm 0.02$  & $0.29\pm 0.07$ \\ 
\midrule[1.6pt] 
$ P_1 (B_s\to \phi\mu^+\mu^-) $ & 2 loop & 1 loop 
& SM~\cite{Descotes-Genon:2015uva} & Expt.~\cite{Aaij:2015esa} \\ 
 \midrule 
 $[0.1,2]$ & $0.013\pm 0.003$ & $0.012\pm 0.002$ 
         & $0.11 \pm 0.08 $ & $ -0.13 \pm 0.33 $ \\ 
 $[2,5]$   & $-0.26\pm 0.05$  & $-0.31\pm 0.06$
         & $-0.10\pm 0.09$  & $-0.38\pm 1.47$  \\ 
 $[5,8]$   & $--$             & $-0.39\pm 0.08$
         & $-0.20\pm 0.10$  & $-0.44\pm 1.27$ \\ 

 $ [11,12.5] $ & $-0.50\pm 0.10$&$-0.50\pm 0.10$
 				& $--$ & $--$  \\ 
 $ [15,17] $ & $-0.71\pm 0.14$&$-0.70\pm 0.14$
 				& $--$ & $--$  \\ 
 $ [17,19] $ & $-0.86\pm 0.17$&$-0.86\pm 0.17$
 				& $--$ & $--$  \\ 
 $ [1,6] $ & $-0.22\pm 0.04$&$-0.28\pm 0.06$
 				& $--$ & $--$  \\ 
 $[15,19]$ & $-0.77\pm 0.15$  & $-0.77\pm 0.15$  
         & $-0.69\pm 0.03$  & $-0.25\pm 0.34$ \\ 
\midrule[1.6pt] 
$ P'_4 (B_s\to \phi\mu^+\mu^-) $ & 2 loop & 1 loop 
& SM~\cite{Descotes-Genon:2015uva} & Expt.~\cite{Aaij:2015esa}\\ 
 \midrule 
 $[0.1,2]$ & $-0.18\pm 0.04$  & $-0.15\pm 0.03$ 
         & $-0.28\pm 0.14$  & $-1.35\pm 1.46$ \\ 
 $[2,5]$   & $0.86\pm 0.17$   & $0.96\pm 0.19$
         & $0.80\pm 0.11$   & $2.02\pm 1.84$  \\ 
 $[5,8]$   & $--$ & $1.15\pm 0.23$
         & $1.06\pm 0.06$   & $0.40\pm 0.72$  \\ 
 $ [11,12.5] $ & $1.22\pm 0.24$ &$1.22\pm 0.24$
 				& $--$ & $--$  \\ 
 $ [15,17] $ & $1.31\pm 0.26$&$1.30\pm 0.26$
 				& $--$ & $--$ \\ 
 $ [17,19] $ & $1.36\pm 0.27$&$1.36\pm 0.27$
 				& $--$ & $--$  \\ 
 $ [1,6] $ & $0.75\pm 0.15$&$0.86\pm 0.17$
 				& $--$ & $--$  \\ 
 $[15,19]$ & $1.33\pm 0.26$   & $1.33\pm 0.26$ 
         & $1.30\pm 0.01$   & $0.62\pm 0.49$  \\ 
\midrule[1.6pt] 
$ P'_6 (B_s\to \phi\mu^+\mu^-) $ & 2 loop & 1 loop 
& SM~\cite{Descotes-Genon:2015uva} & Expt.~\cite{Aaij:2015esa}\\ 
 \midrule 
 $[0.1,2]$ & $-0.016\pm 0.003$ & $0$
         & $-0.06 \pm 0.02 $ & $-0.10\pm 0.30$ \\ 
 $[2,5]$   & $-0.015\pm 0.003$ & $0$
         & $-0.05 \pm 0.02 $ & $0.06\pm 0.49$  \\ 
 $[5,8]$   & $ --$              & $0$
         & $-0.02\pm 0.01$   & $-0.08\pm 0.40$ \\ 
 $ [11,12.5] $ & $-0.0043\pm 0.0008$&$0$
 				& $--$ & $--$ \\ 
 $ [15,17] $ & $-0.0018\pm 0.0004$&$0$
 				& $--$ & $--$ \\ 
 $ [17,19] $ & $-0.00071 \pm 0.00014$&$0$
 				& $--$ & $--$  \\ 
 $ [1,6] $ & $ -0.014\pm 0.003$& $0$
 				& $--$ & $--$ \\ 
 $[15,19]$ & $-0.0014\pm 0.0003$              & $0$ 
         & $-0.00 \pm 0.07$  & $-0.29\pm 0.24$ \\ 
 \midrule[1.6pt] 
$ S_3 (B_s\to \phi\mu^+\mu^-) $ & 2 loop & 1 loop 
& SM~\cite{Descotes-Genon:2015uva} & Expt.~\cite{Aaij:2015esa} \\ 
 \midrule 
 $[0.1,2]$ & $0.0031\pm 0.0006$  & $0.0023\pm 0.0005$ 
         & $0.02  \pm 0.02  $  & $-0.05 \pm 0.13  $  \\ 
 $[2,5]$   & $-0.035\pm 0.007$   & $-0.039\pm 0.008$ 
         & $-0.01 \pm 0.01 $   & $-0.06 \pm 0.21 $   \\ 
 $[5,8]$   & $--$                & $-0.082\pm 0.016$
         & $-0.03\pm 0.02$     & $-0.10 \pm 0.25 $   \\ 
 $ [11,12.5] $ & $-0.15\pm 0.03$&$-0.15\pm 0.03$
  				& $--$ & $ -0.19 \pm 0.21 $  \\ 
 $ [15,17] $ & $-0.23\pm 0.05$&$-0.23\pm 0.05$
 				& $--$ & $ -0.06 \pm 0.18 $ \\ 
 $ [17,19] $ & $-0.29\pm 0.06$&$-0.29\pm 0.06$
 				& $--$ & $ -0.07 \pm 0.25 $ \\ 
 $ [1,6] $ & $ -0.034\pm 0.007$& $-0.039\pm 0.008$
 				& $--$ & $ -0.02 \pm 0.13 $\\ 
 $[15,19]$ & $-0.25\pm 0.05$     & $-0.25\pm 0.05$ 
         & $-0.22\pm 0.01$     & $-0.09\pm 0.12$     \\ 
\midrule[1.6pt] 
$ S_4 (B_s\to \phi\mu^+\mu^-) $ & 2 loop &  1 loop 
& SM~\cite{Descotes-Genon:2015uva} & Expt.~\cite{Aaij:2015esa}\\ 
 \midrule 
 $[0.1,2]$ & $-0.038\pm 0.008$   & $-0.031\pm 0.006$ 
         & $-0.06 \pm 0.03 $   & $-0.27 \pm 0.23 $ \\ 
 $[2,5]$   & $0.19\pm 0.04$      & $0.21\pm 0.04$
         & $0.16\pm 0.03$      & $0.47\pm 0.37$  \\ 
 $[5,8]$   & $--$                & $0.28\pm 0.06$
         & $0.25\pm 0.02$      & $0.10\pm 0.17$  \\ 
 $ [11,12.5] $ & $0.30\pm 0.06$&$0.30\pm 0.06$
 				& $--$ & $ 0.47 \pm 0.25 $ \\ 
 $ [15,17] $ & $0.31\pm 0.06$&$0.31\pm 0.06$
 				& $--$ & $ 0.03 \pm 0.15 $\\ 
 $ [17,19] $ & $0.32\pm 0.06$&$0.32\pm 0.06$
 				& $--$ & $ 0.39 \pm 0.3 $\\ 
 $ [1,6] $ & $0.17\pm 0.03$&$0.19\pm 0.04$
 				& $--$ & $ 0.19 \pm 0.14 $ \\ 
 $[15,19]$ & $0.31\pm 0.06$      & $0.31\pm 0.06$ 
         & $0.31\pm 0.00$      & $0.14\pm 0.11$  \\ 
\midrule[1.6pt] 
$ S_7 (B_s\to \phi\mu^+\mu^-) $ & 2 loop &  1 loop 
& SM~\cite{Descotes-Genon:2015uva} & Expt.~\cite{Aaij:2015esa} \\ 
 \midrule 
 $[0.1,2]$ & $0.0065\pm 0.0013$  &  $0$  
         & $0.03  \pm 0.01  $  &  $0.04\pm 0.12$   \\ 
 $[2,5]$   & $0.0065\pm 0.0013$  &  $0$
         & $0.02  \pm 0.01  $  &  $-0.03\pm 0.21$  \\ 
 $[5,8]$   & $--$  &  $0$
         & $0.01\pm 0.00$      &  $0.04\pm 0.18$ \\ 
 $ [11,12.5] $ & $ 0.0021\pm 0.0004$&$0$
 				& $--$ & $ 0.00 \pm 0.16 $ \\ 
 $ [15,17] $ & $ 0.00087\pm 0.0002$&$0$
 				& $--$ & $ 0.12 \pm 0.15 $  \\ 
 $ [17,19] $ & $ 0.00034\pm 0.00007$&$0$
 				& $--$ & $ 0.20 \pm 0.26 $\\ 
 $ [1,6] $ & $ 0.0065\pm 0.0013$&$0$
 				& $--$ & $ -0.03 \pm 0.14 $  \\ 
 $[15,19]$ & $0.00066\pm 0.00013$ & $0$ 
           & $0.00   \pm 0.03   $ & $0.13\pm 0.11$ \\ 
\bottomrule[1.6pt] 
\end{longtable}

\begin{acknowledgements}

We thank Christoph Greub and Javier Virto for useful discussions
of some aspects of this work.
This work was partly supported by the Slovak Grant Agency for Sciences VEGA, 
grant No. 1/0158/13 (S.~Dubni\v{c}ka, A.Z.~Dubni\v{c}kov\'{a}, A.~Liptaj), 
by the Slovak Research and Development Agency APVV, 
grant No. APVV-0463-12 (S.~Dubni\v{c}ka, A.Z.~Dubni\v{c}kov\'{a}, A.~Liptaj) 
and by Joint research project of Institute of Physics, SAS and 
Bogoliubov Laboratory of Theoretical Physics, JINR, No. 01-3-1114 
(S.~Dubni\v{c}ka, A.Z.~Dubni\v{c}kov\'{a}, M.A.~Ivanov and A.~Liptaj).
A.~Issadykov and M.A.~Ivanov acknowledge the partial support by
the Ministry of Education and Science of the Republic of Kazakhstan,
grant 3092/GF4, state registration~No.~0115RK01040.

\end{acknowledgements}


\begin{thebibliography}{99}

\bibitem{Aaij:2013qta} 
  R.~Aaij {\it et al.} [LHCb Collaboration],
  Phys.\ Rev.\ Lett.\  {\bf 111}, 191801 (2013)
  [arXiv:1308.1707 [hep-ex]].

\bibitem{Aaij:2014pli} 
  R.~Aaij {\it et al.} [LHCb Collaboration],
  JHEP {\bf 1406}, 133 (2014)
  [arXiv:1403.8044 [hep-ex]].

\bibitem{Aaij:2015esa} 
  R.~Aaij {\it et al.} [LHCb Collaboration],
  JHEP {\bf 1509}, 179 (2015)
  [arXiv:1506.08777 [hep-ex]].



\bibitem{Descotes-Genon:2015uva} 
  S.~Descotes-Genon, L.~Hofer, J.~Matias and J.~Virto,
  arXiv:1510.04239 [hep-ph].

\bibitem{DescotesGenon:2012zf} 
  S.~Descotes-Genon, J.~Matias, M.~Ramon and J.~Virto,
  JHEP {\bf 1301}, 048 (2013)
  [arXiv:1207.2753 [hep-ph]].

\bibitem{Descotes-Genon:2013vna} 
  S.~Descotes-Genon, T.~Hurth, J.~Matias and J.~Virto,
  JHEP {\bf 1305}, 137 (2013)
  [arXiv:1303.5794 [hep-ph]].

\bibitem{Descotes-Genon:2013wba} 
  S.~Descotes-Genon, J.~Matias and J.~Virto,
  Phys.\ Rev.\ D {\bf 88}, 074002 (2013)
  [arXiv:1307.5683 [hep-ph]].

\bibitem{Altmannshofer:2013foa} 
  W.~Altmannshofer and D.~M.~Straub,
  Eur.\ Phys.\ J.\ C {\bf 73},  2646 (2013)
  [arXiv:1308.1501 [hep-ph]].

\bibitem{Egede:2010zc} 
  U.~Egede, T.~Hurth, J.~Matias, M.~Ramon and W.~Reece,
  JHEP {\bf 1010}, 056 (2010)
  [arXiv:1005.0571 [hep-ph]].

\bibitem{Beneke:2001at} 
  M.~Beneke, T.~Feldmann and D.~Seidel,
  Nucl.\ Phys.\ B {\bf 612}, 25 (2001)
  [hep-ph/0106067].

\bibitem{Beylich:2011aq}
  M.~Beylich, G.~Buchalla and T.~Feldmann,
  Eur.\ Phys.\ J.\ C {\bf 71} (2011) 1635
  [arXiv:1101.5118 [hep-ph]].

\bibitem{Lyon:2014hpa} 
  J.~Lyon and R.~Zwicky,
  arXiv:1406.0566 [hep-ph].

\bibitem{Straub:2015ica} 
  A.~Bharucha, D.~M.~Straub and R.~Zwicky,
  arXiv:1503.05534 [hep-ph].

\bibitem{Khodjamirian:2010vf} 
  A.~Khodjamirian, T.~Mannel, A.~A.~Pivovarov and Y.-M.~Wang,
  JHEP {\bf 1009}, 089 (2010)
  [arXiv:1006.4945 [hep-ph]].

\bibitem{Kang:2013jaa} 
  X.~W.~Kang, B.~Kubis, C.~Hanhart and U.~G.~Meißner,
  Phys.\ Rev.\ D {\bf 89}, 053015 (2014)
  [arXiv:1312.1193 [hep-ph]].

\bibitem{Horgan:2013hoa} 
  R.~R.~Horgan, Z.~Liu, S.~Meinel and M.~Wingate,
  Phys.\ Rev.\ D {\bf 89}, no. 9, 094501 (2014)
  [arXiv:1310.3722 [hep-lat]].

\bibitem{Horgan:2013pva} 
  R.~R.~Horgan, Z.~Liu, S.~Meinel and M.~Wingate,
  Phys.\ Rev.\ Lett.\  {\bf 112}, 212003 (2014)
  [arXiv:1310.3887 [hep-ph]].

\bibitem{Horgan:2015vla} 
  R.~R.~Horgan, Z.~Liu, S.~Meinel and M.~Wingate,
  PoS LATTICE {\bf 2014}, 372 (2015)
  [arXiv:1501.00367 [hep-lat]].

\bibitem{Asatryan:2001zw} 
  H.~H.~Asatryan, H.~M.~Asatrian, C.~Greub and M.~Walker,
  Phys.\ Rev.\ D {\bf 65}, 074004 (2002)
  [hep-ph/0109140].

\bibitem{Greub:2008cy} 
  C.~Greub, V.~Pilipp and C.~Schupbach,
  JHEP {\bf 0812}, 040 (2008)
  [arXiv:0810.4077 [hep-ph]].



\bibitem{Efimov:1988yd} 
  G.~V.~Efimov and M.~A.~Ivanov,
  Int.\ J.\ Mod.\ Phys.\ A {\bf 4}, 2031 (1989).

\bibitem{Efimov:1993ei}
G.~V.~Efimov and M.~A.~Ivanov, 
{\it The Quark Confinement Model of Hadrons}, 
(CRC Press, Boca Raton, 1993). 

\bibitem{Faessler:2002ut} 
A.~Faessler, T.~Gutsche, M.~A.~Ivanov, J.~G.~K\"orner and V.~E.~Lyubovitskij,\\
  Eur.\ Phys.\ J.\ direct C {\bf 4}, 18 (2002)
  [hep-ph/0205287].

\bibitem{Branz:2009cd}
  T.~Branz, A.~Faessler, T.~Gutsche, M.~A.~Ivanov, J.~G.~K\"orner, 
  V.~E.~Lyubovitskij,
  Phys.\ Rev.\  D {\bf 81}, 034010 (2010).
  [arXiv:0912.3710 [hep-ph]].

\bibitem{Weinberg:1962hj}
  S.~Weinberg,
  Phys.\ Rev.\  {\bf 130}, 776 (1963). 

\bibitem{Salam:1962ap}
  A.~Salam,
  Nuovo Cimento\  {\bf 25}, 224 (1962).

\bibitem{Hayashi:1967hk} 
K.~Hayashi, M.~Hirayama, T.~Muta, N.~Seto, and T.~Shirafuji, 
Fortsch.\ Phys.\ {\bf 15}, 625 (1967).

\bibitem{Dubnicka:2015iwg}
  S.~Dubni\v{c}ka, A.~Z.~Dubni\v{c}kov\'{a}, N.~Habyl, M.~A.~Ivanov, A.~Liptaj 
and G.~S.~Nurbakova,
  Few Body Syst.\  {\bf 57}, no. 2, 121 (2016)
  [arXiv:1511.04887 [hep-ph]].

\bibitem{Ivanov:2011aa} 
  M.~A.~Ivanov, J.~G.~K\"orner, S.~G.~Kovalenko, P.~Santorelli and 
G.~G.~Saidullaeva,
  Phys.\ Rev.\ D {\bf 85}, 034004 (2012)
  [arXiv:1112.3536 [hep-ph]].

\bibitem{Khodjamirian:2006st} 
  A.~Khodjamirian, T.~Mannel and N.~Offen,
  Phys.\ Rev.\ D {\bf 75}, 054013 (2007)
  [hep-ph/0611193].

\bibitem{Ball:2004rg} 
  P.~Ball and R.~Zwicky,
  Phys.\ Rev.\ D {\bf 71}, 014029 (2005)
  [hep-ph/0412079].

\bibitem{Faustov:2013pca} 
  R.~N.~Faustov and V.~O.~Galkin,
  Eur.\ Phys.\ J.\ C {\bf 73}, no. 10, 2593 (2013)
  [arXiv:1309.2160 [hep-ph]].

\bibitem{Yilmaz:2008pa}
  U.~O.~Yilmaz,
  Eur.\ Phys.\ J.\ C {\bf 58} (2008) 555
  [arXiv:0806.0269 [hep-ph]].

\bibitem{Ali:2007ff}
  A.~Ali, G.~Kramer, Y.~Li, C.~D.~Lu, Y.~L.~Shen, W.~Wang and Y.~M.~Wang,
  Phys.\ Rev.\ D {\bf 76} (2007) 074018
  [hep-ph/0703162 [HEP-PH]].

\bibitem{Melikhov:2000yu}
  D.~Melikhov and B.~Stech,
  Phys.\ Rev.\ D {\bf 62} (2000) 014006
  [hep-ph/0001113].

\bibitem{Li:2009tx}
  R.~H.~Li, C.~D.~Lu and W.~Wang,
  Phys.\ Rev.\ D {\bf 79} (2009) 034014
  [arXiv:0901.0307 [hep-ph]].

\bibitem{Lu:2007sg}
  C.~D.~Lu, W.~Wang and Z.~T.~Wei,
  Phys.\ Rev.\ D {\bf 76} (2007) 014013
  [hep-ph/0701265 [HEP-PH]].

\bibitem{Wu:2006rd}
  Y.~L.~Wu, M.~Zhong and Y.~B.~Zuo,
  Int.\ J.\ Mod.\ Phys.\ A {\bf 21} (2006) 6125
  [hep-ph/0604007].

\bibitem{Buchalla:1995vs} 
  G.~Buchalla, A.~J.~Buras and M.~E.~Lautenbacher,
  Rev.\ Mod.\ Phys.\  {\bf 68}, 1125 (1996)
  [hep-ph/9512380].

\bibitem{Ali:1991is}
A. Ali, T. Mannel, T. Morozumi: 
Phys. \ Lett. B {\bf 273} 505 (1991). 

\bibitem{Bobeth:2008ij} 
  C.~Bobeth, G.~Hiller and G.~Piranishvili,
  JHEP {\bf 0807}, 106 (2008)
  [arXiv:0805.2525 [hep-ph]].

\bibitem{Descotes-Genon:2015hea} 
  S.~Descotes-Genon and J.~Virto,
  JHEP {\bf 1504}, 045 (2015)
  [JHEP {\bf 1507}, 049 (2015)]
  [arXiv:1502.05509 [hep-ph]].

\bibitem{Geng:2003su}
  C.~Q.~Geng and C.~C.~Liu,
  J.\ Phys.\ G {\bf 29} (2003) 1103
  [hep-ph/0303246].

\bibitem{Agashe:2014kda} 
  K.~A.~Olive {\it et al.} [Particle Data Group Collaboration],
  Chin.\ Phys.\ C {\bf 38}, 090001 (2014).


\end{thebibliography}
\end{document}